\documentclass[sigconf]{acmart}

\copyrightyear{2023} 
\acmYear{2023}  
\setcopyright{acmcopyright}
\acmConference[WSDM '23]{Proceedings of the Sixteenth ACM
International Conference on Web Search and Data Mining}{February  27-March     3,
2023}{Singapore, Singapore}
\acmBooktitle{Proceedings of the Sixteenth ACM International Conference on Web
Search and Data Mining (WSDM '23), February  27-March     3, 2023, Singapore,
Singapore}
\acmPrice{15.00}
\acmDOI{10.1145/3539597.3570445}
\acmISBN{978-1-4503-9407-9/23/02}

\usepackage[doipre={doi:~}]{uri}
\AtBeginDocument{%
  \providecommand\BibTeX{{%
    \normalfont B\kern-0.5em{\scshape i\kern-0.25em b}\kern-0.8em\TeX}}}

\usepackage{tabularx}
\usepackage{subfigure}
\usepackage{multicol}
\usepackage{multirow}
\usepackage{balance}
\usepackage{algorithm}
\usepackage{algorithmicx}
\usepackage[T1]{fontenc}
\usepackage{aecompl}
\usepackage{algpseudocode}

\begin{document}

\title{Efficiently Leveraging Multi-level User Intent for Session-based Recommendation via Atten-Mixer Network}

\author{Peiyan Zhang$^{*}$}\thanks{$^{*}$Equal contribution.}
\affiliation{\institution{The Hong Kong University of \\ Science and Technology}\country{Hong Kong}}
\email{pzhangao@cse.ust.hk}

\author{Jiayan Guo$^{*}$}

\affiliation{\institution{School of Intelligence Science and Technology, Peking University}\city{Beijing}\country{China}}
\email{guojiayan@pku.edu.cn}

\author{Chaozhuo Li$^{*}$}
\affiliation{\institution{Microsoft Research Asia}\city{Beijing}\country{China}}
\email{cli@microsoft.com}

\author{Yueqi Xie}
\affiliation{\institution{The Hong Kong University of \\ Science and Technology}\country{Hong Kong}}
\email{yxieay@cse.ust.hk}

\author{Jae Boum Kim}
\affiliation{\institution{The Hong Kong University of \\ Science and Technology}\country{Hong Kong}}
\affiliation{\institution{Upstage}\country{Republic of Korea}}
\email{jbkim@cse.ust.hk}

\author{Yan Zhang}
\affiliation{\institution{School of Intelligence Science and Technology, Peking University}\city{Beijing}\country{China}}
\email{zhyzhy001@pku.edu.cn}

\author{Xing Xie}
\affiliation{\institution{Microsoft Research Asia}\city{Beijing}\country{China}}
\email{xing.xie@microsoft.com}

\author{Haohan Wang}
\affiliation{\institution{University of Illinois at Urbana-Champaign}\city{IL}\country{USA}}
\email{haohanw@illinois.edu}

\author{Sunghun Kim}
\affiliation{\institution{The Hong Kong University of \\ Science and Technology}\country{Hong Kong}}
\email{hunkim@cse.ust.hk}

\renewcommand{\shortauthors}{Peiyan Zhang, et al.}

\begin{abstract}
  Session-based recommendation (SBR) aims to predict the user’s next action based on short and dynamic sessions. Recently, there has been an increasing interest in utilizing various elaborately designed graph neural networks (GNNs) to capture the
pair-wise relationships among items,
seemingly suggesting the design of more complicated models 
is the panacea for improving the empirical performance. 
However, these models achieve relatively marginal improvements with exponential growth in model complexity.
In this paper, we dissect the classical GNN-based SBR models and empirically find that some sophisticated GNN propagations are redundant, given the readout module plays a significant role in GNN-based models. Based on this observation, we intuitively propose to remove the GNN propagation part, while the readout module will take on more responsibility in the model reasoning process. To this end, we propose the Multi-Level Attention Mixture Network (Atten-Mixer), which leverages both concept-view and instance-view readouts to achieve multi-level reasoning over item transitions. As simply enumerating all possible high-level concepts is infeasible for large real-world recommender systems, we further incorporate SBR-related inductive biases, \textit{i.e.,} local invariance and inherent priority to prune the search space. Experiments on three benchmarks demonstrate the effectiveness and efficiency of our proposal. We also have already launched the proposed techniques to a large-scale e-commercial online service since April 2021, with significant improvements of top-tier business metrics demonstrated in the 
online experiments on live traffic. Our code is available
at \href{https://github.com/Peiyance/Atten-Mixer-torch}{https://github.com/Peiyance/Atten-Mixer-torch}.
\end{abstract}






\maketitle


\section{Introduction}
\label{sec:intro}
Recommender systems play vital roles on various online platforms, due to their success in addressing information overload challenges by recommending useful content to users \cite{xie2022decoupled,zhou2022equivariant}. Conventional recommendation approaches (e.g., collaborative filtering~\cite{CF_early}) usually rely on the availability of user profiles and long-term historical interactions, and may perform poorly in many recent real-world scenarios, e.g., mobile stream media like YouTube and Tiktok, when traditional collaborative signals are unavailable (e.g., unlogged-in user) or limited (e.g., short-term historical interaction)~\cite{gao2022graph,Hidasi2016SessionbasedRW}. Consequently, session-based recommendation has attracted extensive attention, which predicts the user's next action based on sessions containing limited behavioral information.

Recent SBR research sees a proliferation of usages of GNN-based models to better capture the complex transitions of items. Wu et al.~\cite{Wu2019SessionbasedRW} first propose to capture the pair-wise relations with a simple Graph Gated Neural Network~\cite{li2015gated}. Afterward, Pan et al.~\cite{Pan2020StarGN} construct star graphs and add highway networks~\cite{Srivastava2015HighwayN} to avoid overfitting. Beyond pair-wise relations, Xia et al.~\cite{xia2021self} propose a dual channel hypergraph convolutional network to consider the high order information among items. However, compared with the exponential growth in model complexity, the performance gain on benchmarks brought by each model is marginal~(see Table~\ref{tab:overall} for more details). In view of this phenomenon, a meaningful question naturally arises:~\textit{Are those GNN-based models under- or over- complicated for SBR?} To answer this question, we dissect the existing GNN-based SBR models and empirically find that some GNN propagations seem redundant, given the readout module plays a significant role in these models (see Section~\ref{sec:dissect} for more details).

This observation is quite counter-intuitive to today's tendency where the SBR community seeks more powerful GNN designs to capture the complex transitions among items~\cite{Wu2019SessionbasedRW,Yu2020TAGNNTA,Wang2020GlobalCE,Wang2020PAGGANSR}. 
Compared with other recommendation areas, the session graph is far more sparse due to the intrinsic short and dynamic properties of session data. For example, almost 70\% of sessions in \textit{Diginetica} dataset are composed of distinct items, which means that constructing the graph based on the session data may merely result in a sequence. 
In this case, some designs in GNNs are rather heavy and burdensome, only contributing marginally compared to the overall preference that the readout module has managed to learn from the data. 
Therefore, we hypothesize an advanced architecture design on the readout module will benefit more. As we loose the requirement for GNN propagation part, the readout module should take on more responsibility in the model reasoning process. Thus, a readout module with powerful reasoning ability is desired.

\begin{figure}[t]
    \centering
    \includegraphics[width=.9\linewidth]{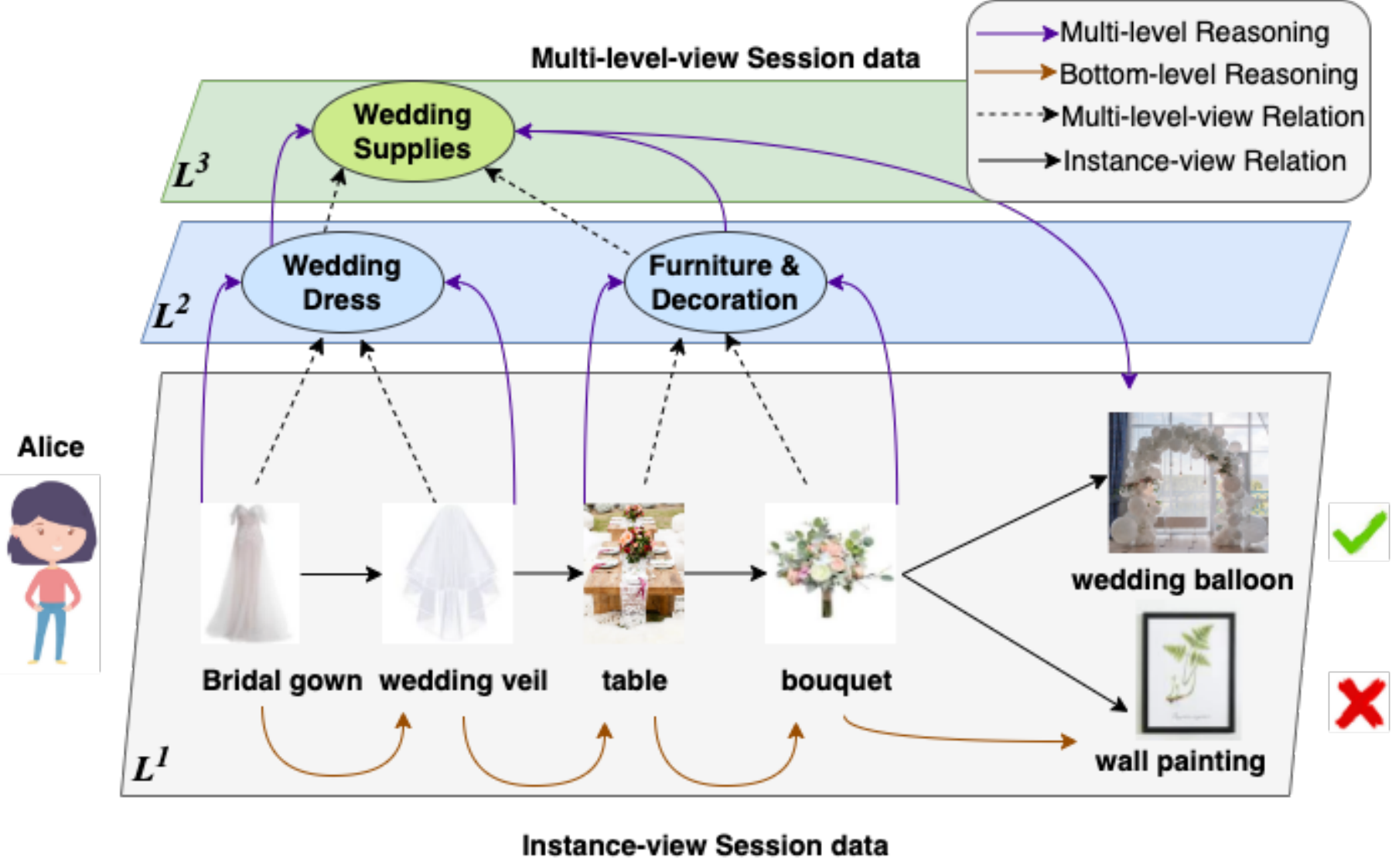}
    \caption{An example of multi-level reasoning over session data.}
    \label{fig:toy}
\end{figure}

Existing works on recommendation readout modules mostly focus on instance-view readout~\cite{wang2021survey}. As shown in Figure~\ref{fig:toy}, an instance-view
readout module generates the overall preference with attention mechanism related to the single item, \textit{i.e.,} every item~\cite{Pan2020StarGN, Li2017NeuralAS, Wang2020PAGGANSR, Wang2020GlobalCE,Xu2019GraphCS} or last-clicked item~\cite{Wu2019SessionbasedRW,chen2020handling,Yu2020TAGNNTA}. Without the reasoning process brought by the GNN propagation part, instance-view readout lacks information about high-level connections between items, \textit{i.e.,} both bridal gown and wedding veil are wedding supplies. Based on instance-view readout module, we can only perform bottom-level reasoning over item transitions. However, bottom-level reasoning process is fragmented since it cannot directly correlate with different user behaviors. As shown in Figure~\ref{fig:toy}, if we correlate different
behaviors of Alice by considering the high-level connections between items, we can see that Alice tends to purchase
balloons of wedding supplies rather than wall paintings of home decoration. However, it is difficult for instance-view readout module to identify this overall pattern, since they only model bottom-level relations. 
Without an accurate understanding of the user behaviors, the
fragmented bottom-level reasoning process has a larger probability
to converge to a local optimum, e.g., recommending an item that
belongs to the home decoration to Alice. 

To address this issue, we take a more comprehensive view of readout modules. Inspired by the ontological knowledge graph (KG) reasoning process~\cite{hao2019universal,wang2022multi} which considers relating higher-level concepts over KGs, we propose to combine the high-level-view with instance-view readout together to achieve the multi-level reasoning over item transitions. In particular, designing a multi-level reasoning readout module is challenging, as 
simply enumerating all possible high-level concepts is infeasible for large real-world recommender systems. To address this problem, it is essential to incorporate SBR-related inductive biases, \textit{i.e.,} inherent priority as well as the local invariance~\cite{chen2019session}, which significantly prunes our search space. The inherent priority lies in the emphasis on the contribution of the last few items for the user preference~\cite{choi2021session}, while the local invariance indicates the local order of the last few items is unimportant~\cite{chen2019session}. 

To this end, we propose a Multi-Level Attention Mixture Network (Atten-Mixer), which leverages multi-level user intent to enhance its reasoning ability. Instead of generating one attention map with the query related to the instance-view item, we generate a pool of attention maps based on the multi-level user intent, and then mix them with a simple $L_{p}$ pooling~\cite{hyvarinen2007complex}, which can be viewed as the integration of multi-level user intent. 
The whole model is simple and elegant, which achieves better empirical performance than previous attention-based methods, as well as other state-of-the-art SBR models, even without any enhancement of item embedding using GNNs, within a shorter span of time. In addition, Atten-Mixer can be easily incorporated into other models, and to further help improve their performance. 

Our primary contributions can be summarized as follows:

\begin{itemize}
    \item We analyze the importance of various parts of existing GNN-based SBR models. We empirically find that those GNNs are over-complicated for SBR, while an advanced architecture design on attention-based readout method for the session representation will benefit more.
    \item  We propose a general framework for leveraging multi-level user intent to achieve multi-level reasoning over item transitions, which can be easily integrated into existing SBR models to further boost the performance. We incorporate several SBR-related inductive biases to prune the search space and ensure a well balance between efficiency and recommendation accuracy.
    \item Our methods have been fully deployed into our live system as the default choice to serve millions of page views each day, and
consistently outperform the previous state-or-art baselines. Offline analyses on three benchmark datasets are provided towards the rationality of Atten-Mixer from both technical and empirical perspectives.

\end{itemize}


\section{Related Works}
Instead of utilizing user profiles and historical actions, session-based recommender systems learn to model users' preferences and recommend the next interaction solely based on short and dynamic sessions. 
In this section, we review the recent deep learning-based works on SBR.



Sharing some common sequential characteristics with neural language processing~\cite{batmaz2019review}, research on session-based recommendation takes advantage of the rapid development of language models. Hidasi et al.~\cite{Hidasi2016SessionbasedRW} first propose to leverage the recurrent neural networks (RNNs) to model users' preferences.
Afterward, attention-based mechanisms~\cite{vaswani2017attention} are incorporated into the system and significantly boost performance. NARM~\cite{Li2017NeuralAS} utilizes attention on RNN models to enhance features while STAMP~\cite{Liu2018STAMP} captures long and short-term preferences relying on a simple attentive model.

Convolution Neural Networks (CNNs) are also leveraged in session-based recommendation. Tang et al.~\cite{tang2018personalized} try to embed item session as a matrix and perform convolution on the matrix to get the representation. 
Except for the usage of CNNs only, Hidasi et al.~\cite{Hidasi2016SessionbasedRW} incorporate CNNs with RNNs to take advantage of both long and short-term dependencies. 
These deep learning-based models greatly improve the prediction accuracy with a strong capacity in modeling the complex session. 

To better model the transitions within the sessions, most recent developments focus on leveraging Graph Neural Networks (GNNs) to extract the relationship \cite{li2021adsgnn,huang2022going,guo2022evolutionary}. Wu et al.~\cite{Wu2019SessionbasedRW} first propose to capture the complex transitions with graph structure. Afterward, Pan et al.~\cite{Pan2020StarGN} try to avoid overfitting through highway networks~\cite{Srivastava2015HighwayN}. Position information~\cite{Wang2020PAGGANSR}, target information~\cite{Yu2020TAGNNTA}, and global context~\cite{Wang2020GlobalCE} are also taken into consideration to further improve the performance. 
Some recent works~\cite{liu2020long,gupta2021causer,gupta2019niser} pay attention to the issue of popularity bias~\cite{zheng2021disentangling} and information loss~\cite{chen2020handling}. By solving these problems, the GNN-based models are further enhanced.
As outlined above, previous efforts emphasize more the GNN propagation part, while few efforts are devoted to designing effective readout operations to aggregate these embeddings to the session-level embedding. 
Moreover, as discussed in Section~\ref{sec:intro}, the current readout operations have limited capacity in reasoning over sessions. Also, the performance improvement of GNN models is undesirable compared with the time-consuming and memory consumption brought by sophisticated GNN models. 

\section{Analysis on GNN-based SBR models}
\label{sec:dissect}
In this section, we first formulate the task of SBR and sketch out the general workflow of standard GNN-based SBR models (Section~\ref{sec:problem_definition}). Afterward, we illustrate the process that applying SparseVD to automatically dissect the GNN architecture (Section~\ref{sec:svd}). Finally, we analyze the results to conclude our observations (Section~\ref{sec:analysis}). 

\subsection{Preliminaries}
\label{sec:problem_definition}
Session-based recommendation is a special case of next-item prediction. 
It is to predict the next-item given the current active session 
containing a sequence of clicked items. Assume the item set is $V=\{v_1, v_2,..., v_{|V|}\}$, where $v_{j}$ indicates item $j$ and $|V|$ denotes the number of all items. Given an ongoing session denoted as $S=\{v_1, v_2,..., v_{n}\}$, the aim of a session-based recommendation is to predict the item that the user will interact with at the next timestamp, that is, $v_{n+1}$.
Typical session-based recommendation takes the 
input $S$ to generate probability distributions of the next item $\hat{y} = p(v_{n+1}|S)$ with 
each entry denotes the relevance score corresponding to the items in $V$. Then the items with the top-K scores are used to make a recommendation.

A wealth of existing works~\cite{Wu2019SessionbasedRW,Qiu2019RethinkingTI,wang2020global} have explored to leverage GNNs to model the transitions within the sessions. These methods usually have complex architecture designs to process complex pair-wise item relationships, which typically contain two modules: (1) multiple GNN layers to propagate the pair-wise transition patterns along the edges, and (2) an attention-based readout component that aggregates items within the session to compute a compact session representation.

\subsection{Empirical Explorations}
\label{sec:svd}
\begin{figure}[h]
    \centering
    \includegraphics[width=.8\linewidth]{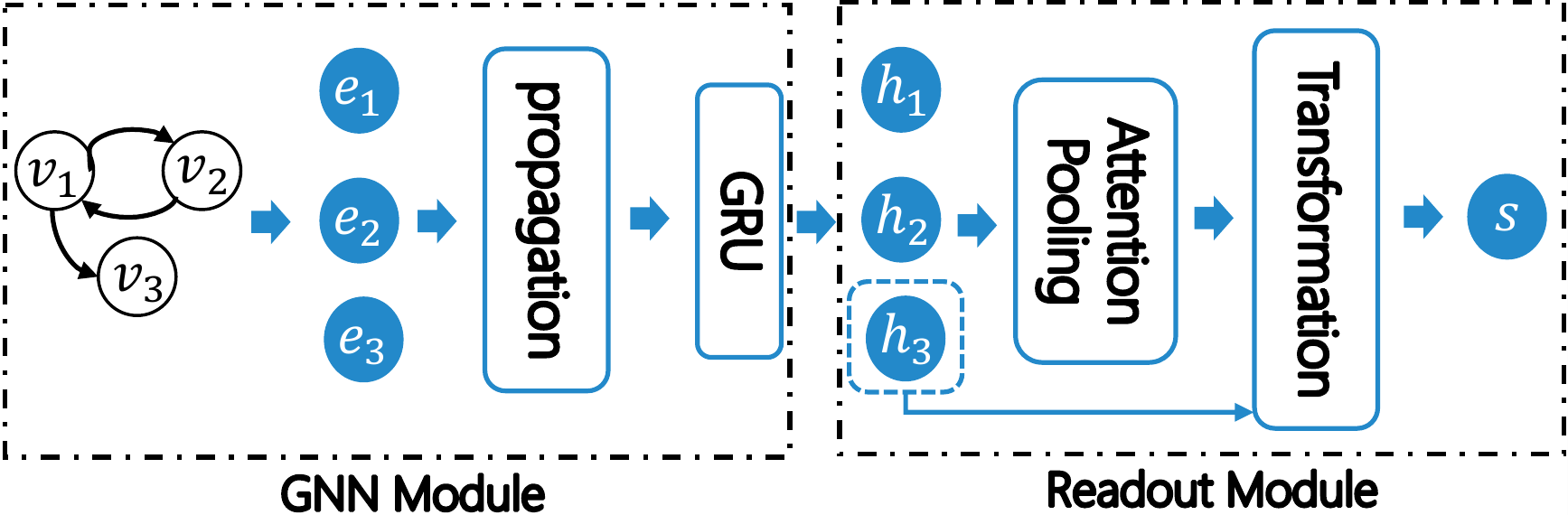}
    \caption{Decomposition of typical GNN-based SBR models.}
    \label{fig:srgnn_decomp}
\end{figure}

To empirically understand the inner mechanism of session-based recommendation, we firstly decompose the typical GNN-based SBR models into two parts, the GNN module and the Readout module. The parameters of each module are shown in Figure \ref{fig:srgnn_decomp}.

\begin{figure}[t]
\centering
\subfigure[Gowalla]{
\includegraphics[width=.3\linewidth]{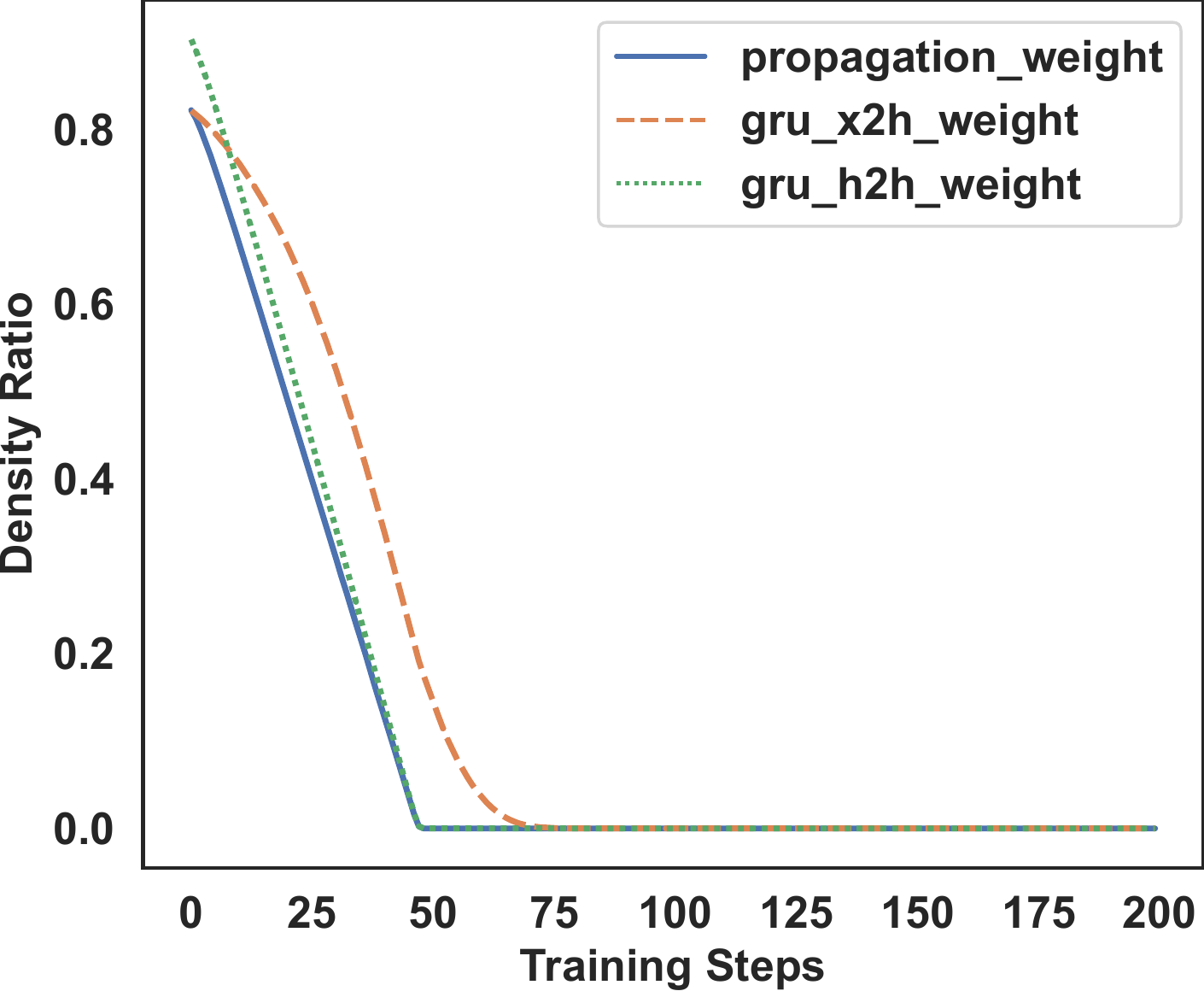}
\label{fig:sparsevd_srgnn}
}
\subfigure[Diginetica]{
\includegraphics[width=.3\linewidth]{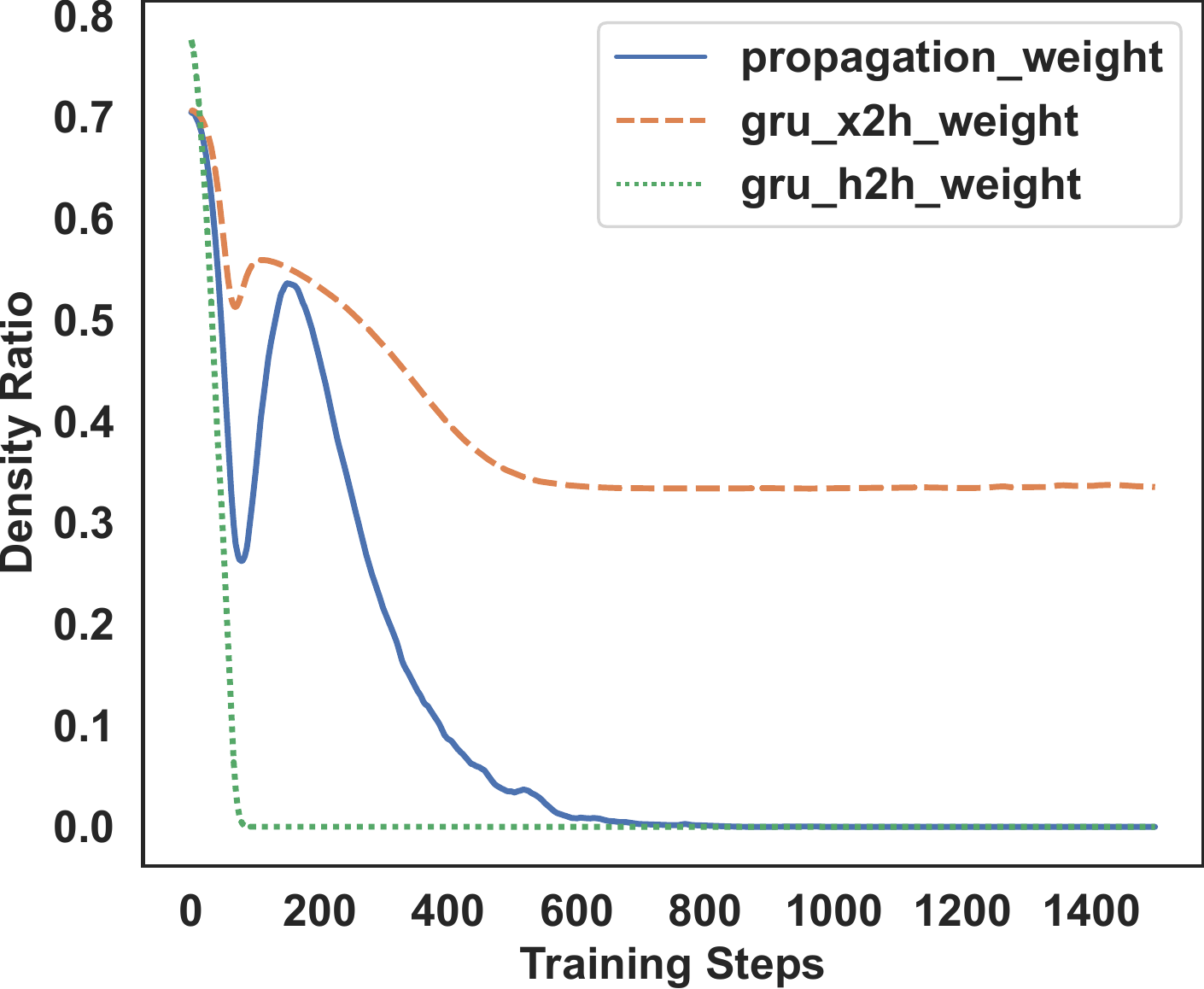}
\label{fig:sparsevd_srgnn}
}
\subfigure[LastFM]{
\includegraphics[width=.3\linewidth]{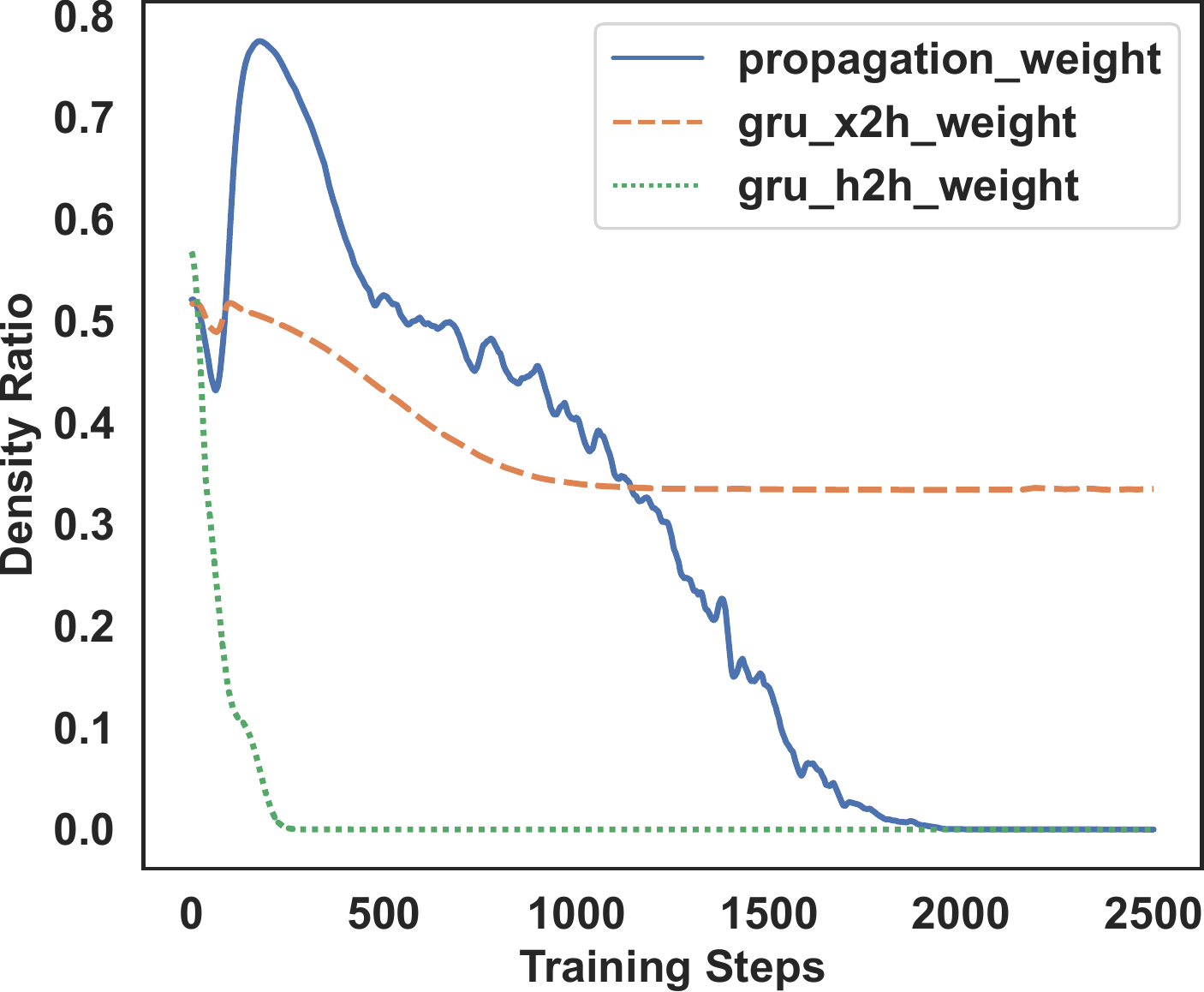}
\label{fig:sparsevd_srgnn}
}

\subfigure[Gowalla]{
\includegraphics[width=.3\linewidth]{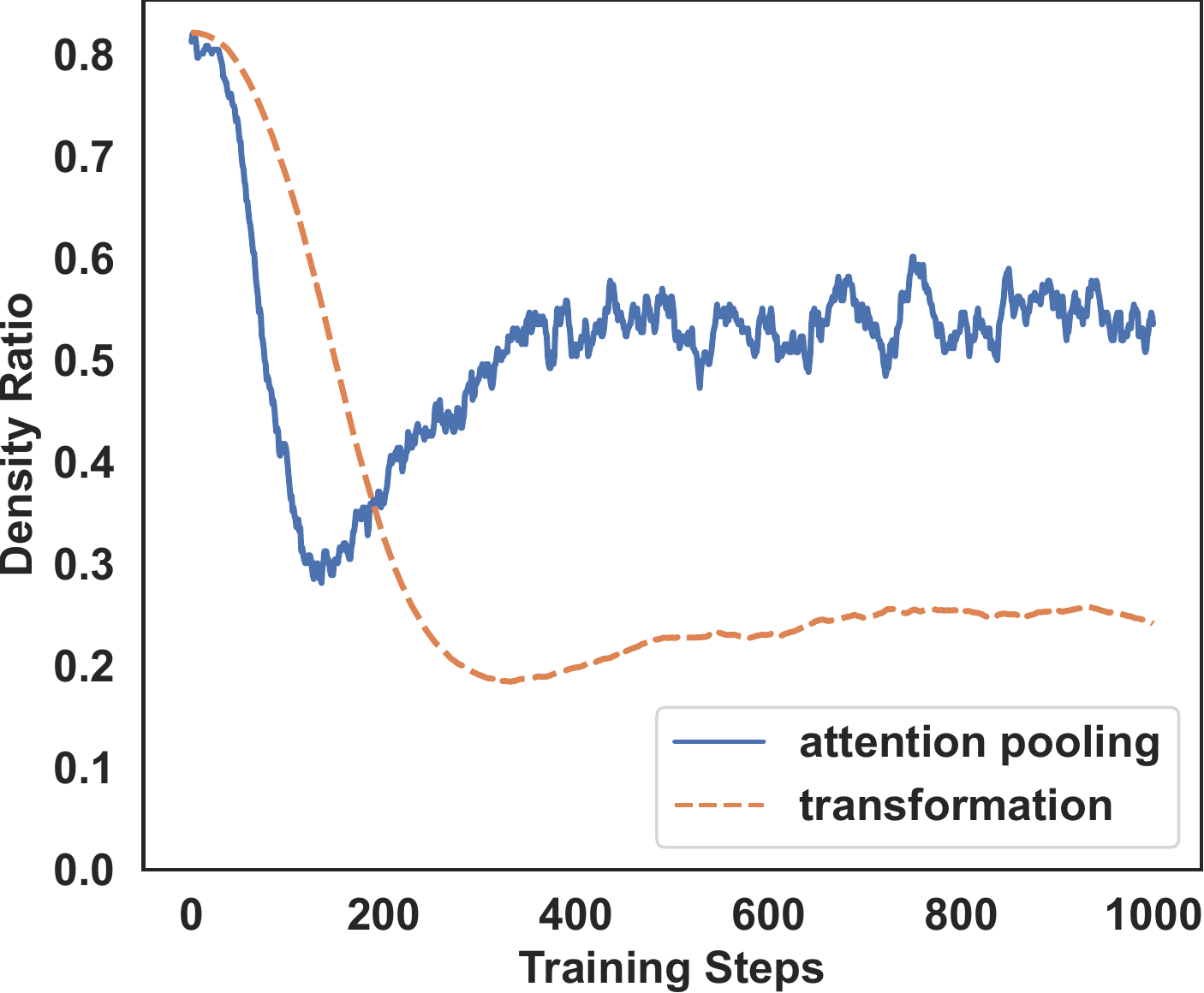}
\label{fig:sparsevd_attn}
}
\subfigure[Diginetica]{
\includegraphics[width=.3\linewidth]{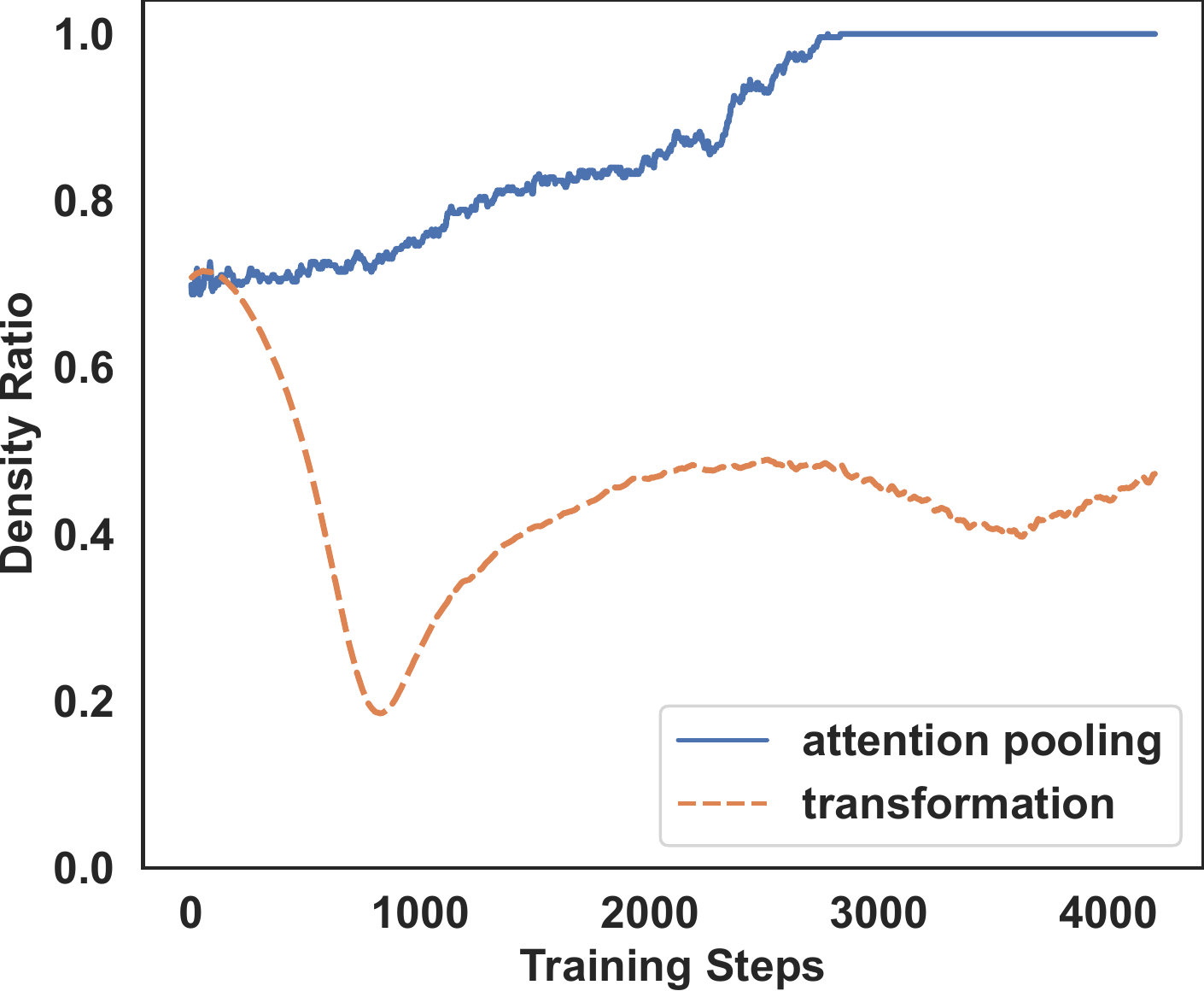}
\label{fig:sparsevd_attn}
}
\subfigure[LastFM]{
\includegraphics[width=.3\linewidth]{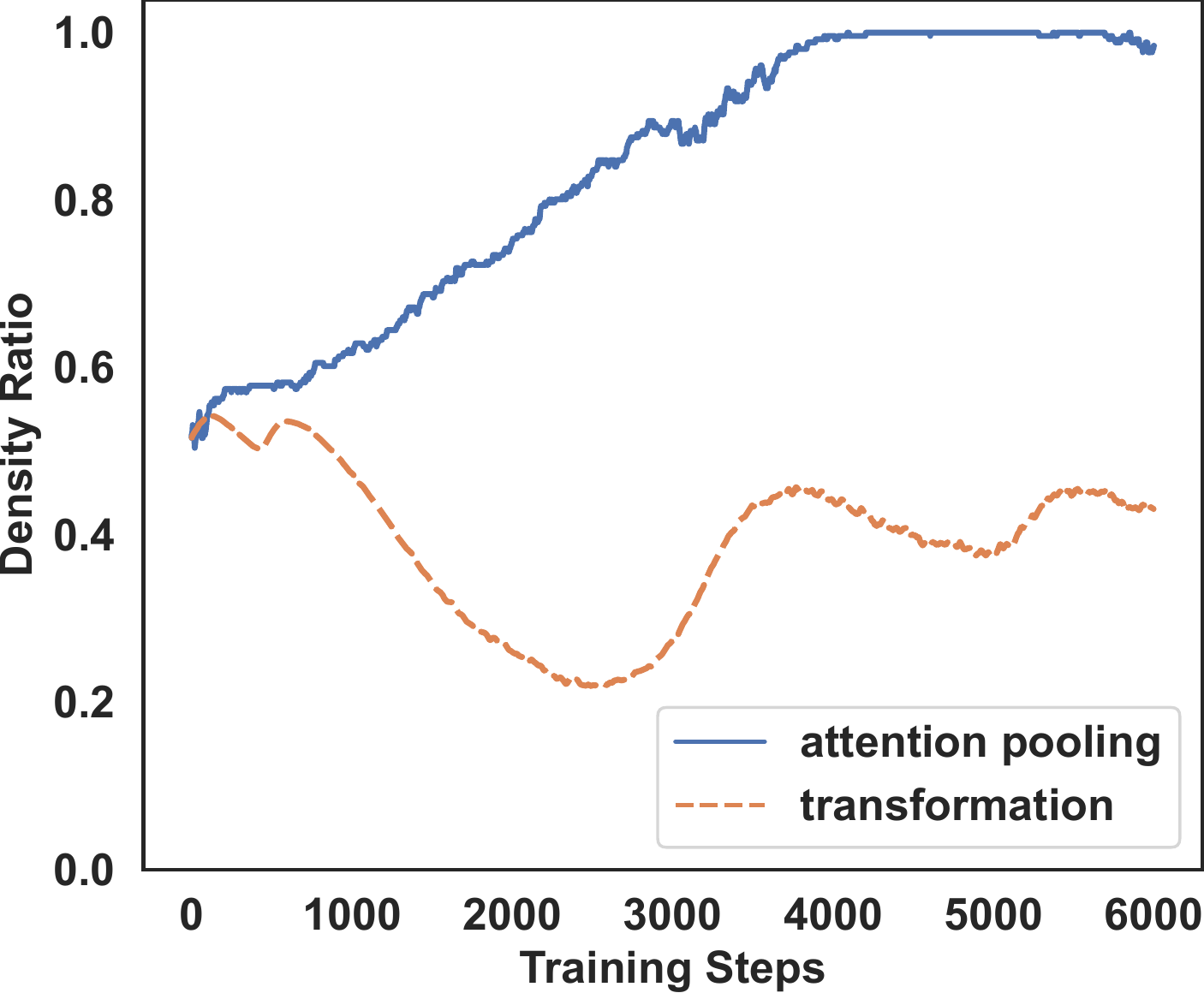}
\label{fig:sparsevd_attn}
}

\caption{Sparsification result by applying SparseVD on GNN Module.} 
\label{fig:svd_gowalla}
\end{figure}
 


For the GNN module, the parameters can be divided into the propagation weights of the graph convolution and GRU weights that combine the original embedding and the output of the graph convolution. While for the Readout module, the parameters are the attention pooling weights to generate long-term representation and transformation weights to generate the session representation for prediction. Then we apply Sparse Variational Dropout~(SparseVD)~\cite{molchanov2017variational}, a widely used technique for neural network sparsification, on the two parts respectively and compute the density ratio while training the model. SparseVD applies Additive Noise Reparameterization~\cite{molchanov2017variational} and Local Reparameterization Trick~\cite{kingma2015variational} to model the distribution of the weights of a neural network:

\begin{equation}
    \begin{split}
        w_{mj}&=\mathcal{N}(\gamma_{mj}, \delta_{mj}) \\
        \gamma_{mj}=\sum_{i=1}^Ia_{mi}&\theta_{ij}, \ \ \delta_{mj}=\sum_{i=1}^Ia_{mi}^2\sigma_{ij}^2 
    \end{split}
\end{equation}

\noindent where $w_{mj}$ is a neuron of position $m,j$ of the weight $W$, $\gamma_{mj}$ and $\delta_{mj}$ are the parameters of a Gaussian distribution which are computed by a mixture of Gaussian distributions $\mathcal{N}(\theta_{ij},\sigma^2_{ij}), \, i=1,...,I$ with the coefficients $a_{mi}, \ ,i=1,...,I$. Then it adds a negative KL-divergence regularization term to the loss function to force the weights to be sparse while keeping the performance:

\begin{equation}
    \mathcal{L}_{reg}=-D_{KL}\left (q(w_{ij}|\theta_{ij}, \alpha_{ij})||p(w_{ij})\right )
    \label{eq:reg}
\end{equation}

\noindent where $p(w_{ij})$ is a standard normal distribution. By applying SparseVD, the regularization term (Eq.~\ref{eq:reg}) will force the mean $\gamma_{mj}$ and variance $\delta_{mj}$ of the weight $w_{mj}$ to approach to a standard normal distribution, thus some unimportant weights are approaching to 0 and only important weights will be kept during the training procedure. Then we compute the density ratio of each weight, which is the ratio of entries of the weight larger than a threshold value $\alpha$:

\begin{equation}
    \rho_{\text{density}}=\frac{1}{MN}\sum_{i=1}^M\sum_{j=1}^{N}I(w_{ij}>\alpha)
\end{equation}

\noindent where $M$ and $N$ are the number of rows and columns of a weight matrix and $w_{ij}$ is the value under index $i,j$. $I(\cdot)$ is an indicator function, where it equals to 1 if the condition is true otherwise 0. 

\begin{figure*}[h]
    \centering
    \includegraphics[width=.8\linewidth]{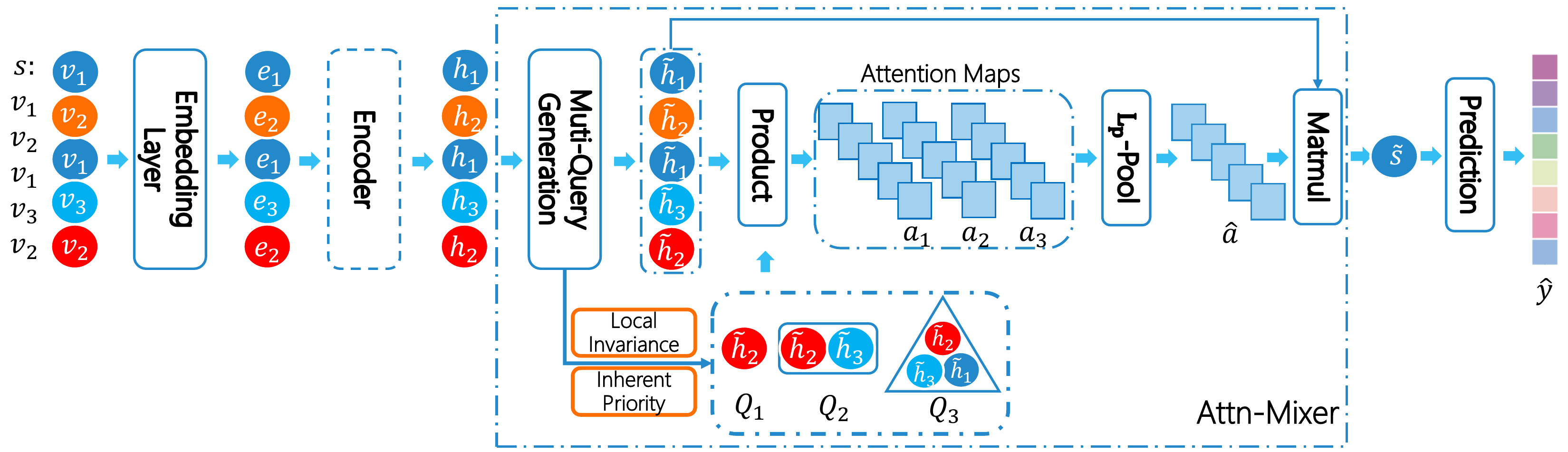}
    \caption{Overview of the Atten-Mixer. Given a session, we first gather normalized item embeddings from the embedding layer and apply Multi-head Level-L Attention (L = 3) by using deep sets operation on the last $l$ normalized hidden states. Then the $L_{p}$ pooling is used on the generated attention maps to get session representation and make recommendation. Our framework supports employing various item encoders~(doted square) before Attn-Mixer.}
    \label{fig:architecture}
\end{figure*}

\subsection{Observations}
\label{sec:analysis}
According to these plots, we summarize our key observations as follows: (1) In GNN module, the density ratio of graph propagation weights is approaching zero as training progresses, indicating the over-complicated GNN design in SBR. (2) In Readout module, the attention pooling weight could preserve a relatively higher density ratio. We observe similar results in other SBR models.

Inspired by these findings, we come up with the following design guidelines for a much simpler yet effective model for SBR: (1) instead of emphasizing the complex GNN design, we tend to remove the GNN propagation part and only preserve the initial embedding layer; (2) we should focus more on the attention-based readout module. As the attention pooling weight preserves the highest density ratio, we hypothesize an 
advanced architecture design on attention-based readout method for the session representation will benefit more. As we loose the requirement for GNN propagation part, the readout module should take on the responsibility in the model reasoning process. Considering the insufficient reasoning ability of existing instance-view readout modules, a readout module with a more powerful reasoning ability is desired.

According to psychopathology, human reasoning is by nature a multi-level information processing procedure~\cite{heriot2012multi}. For example, humans may first consider Alice’s interest with respect to high-level concepts, \textit{e.g.,} whether she is intended to prepare a wedding or decorate a new house. After identifying that Alice is likely to prepare a wedding, humans may then consider wedding supplies related with bouquets, \textit{i.e.,} wedding balloons rather than decoration supplies related with bouquets, \textit{i.e.,} wall paintings. In recommendation reasoning,
adopting such a multi-level reasoning strategy can help prune the large search space, avoid local minimum, and converge to a more satisfying
solution by considering the overall user behaviors. We strive to operationalize this insight into our readout module architecture with a multi-level reasoning component through reflecting on the human reasoning process.

\section{Methodology}

In this section, we illustrate the overall workflow of the proposed Atten-Mixer model (Section~\ref{sec:workflow}). Then, we detail the Atten-Mixer component, involving Multi-level User Intent Attention Generation and Attention Mixture (Section~\ref{sec:attenmixer}).

\subsection{Overall Workflow}
\label{sec:workflow}
The overall workflow is shown in Figure~\ref{fig:architecture}. For an input session $S=\{v_{1},v_{2},...,v_{n}\}$, where $n$ is the length of session $S$,
we gather the embedding of each item from the embedding layer. Each unique item $v_j$ is embedded into a $d$-dimension embedding $h_j$, which is randomly initialized and trainable. 
Then, we normalize the embedding with $\tilde{h}_{j} = \frac{h_{j}}{\left\|h_{j}\right\|_2}$, following the normalization setting in the prior works~\cite{abdollahpouri2017controlling, gupta2019niser}.
Afterward, the proposed Atten-Mixer generates and mixes corresponding attention maps to output the session embedding $\tilde{s}$. Detailed operations are demonstrated in Section~\ref{sec:attenmixer}. It should be noted that we do not elaborate the encoder part shown in Figure~\ref{fig:architecture} as our method supports any kind of encoders that produce item representations. We show the performance of our model with various GNN encoders in Section~\ref{sec:enhanced}.

After obtaining the embedding of the session, we generate the hybrid user preference $\hat{z}_{j}$ and compute the score $\hat{y}_{j}$ for each candidate item $v_j \in V$.
Specifically, we utilize the normalized embedding $\tilde{h}_{j} = \frac{h_{j}}{\left\|h_{j}\right\|_2}$, the normalized session representation $\tilde{s} = \frac{s}{\left\|s\right\|_2}$ and the local preference vector $\tilde{h}_{{n}}$. The steps are as follows:
\begin{equation}
    \hat{z}_{j}=({\rm W}_m(\tilde{s}||\tilde{h}_{{n}}))^{T}\tilde{h}_{j},
\end{equation}
\begin{equation}
    \hat{y}_{j}=\text{softmax}(\sigma \hat{z}_{j}),
\end{equation}
where $||$ denotes the concatenation of two vectors, $\sigma$ is the scaled factor for softmax operation, $\hat{\textbf{y}}=\{\hat{y}_{1},\hat{y}_{2},...,\hat{y}_{|V|}\}$ is the normalized scores vector for all candidate items, and
$W_m \in \mathbb{R}^{d\times 2d}$ is a learnable matrix to transform the concatenation of the session embedding and the local preference. 
\subsection{Atten-Mixer}
\label{sec:attenmixer}
In this section, we elaborate on the details of Atten-Mixer, as shown in Figure~\ref{fig:architecture}. 
Overall, the aim of Atten-Mixer is to serve as an effective readout operation to aggregate the item-level embedding to a representative session embedding. To achieve this goal, the proposed Atten-Mixer mainly involves two steps. The first one is to generate a pool of attention maps with multi-level user intent information based on the inductive biases for the recommendation task. Secondly, we need to reasonably mix these attention maps and perform attention over the sequence of item embedding. 

\subsubsection{Multi-level User Intent Generation}
To generate attention maps with rich semantic information while avoiding disturbance of excessive noisy information, we leverage the inherent priority and the local invariance for SBR~\cite{chen2019session}. The inherent priority lies in the emphasis on the contribution of the recent clicks for the user preference, while the local invariance indicates the local order might not be important. To incorporate these two properties, we first adopt the permutation invariant operation, deep sets~\cite{zaheer2017deep}, on the groups of last items with different lengths to form group representations.
Then we apply linear transformations to these group representations to generate multi-level user intent queries:  

\begin{equation}
\begin{split}
    \mathrm{Q}_{1}&={\mathrm{W}}_{q1}(\tilde{h}_{n}), \\
    {\mathrm{Q}}_{2}&={\mathrm{W}}_{q2}(\Sigma(\{\tilde{h}_{j}\}_{j={n,n-1}})), \\
    ...,&\\
    {\mathrm{Q}}_{L}&={\mathrm{W}}_{qL}(\Sigma(\{\tilde{h}_{j}\}_{j=n,...,n-L+1})), \\
\end{split}
\end{equation}

\noindent where $\tilde{h}_{j}$ is the normalized hidden state of the $j^{th}$ item in the session, and $W_{ql} \in\mathbb{R}^{d\times d}$, $l = 1, 2, ..., L$ refers to a learnable matrix to transform the deep sets results of last $l$ hidden states to an $d$-dimension query ${\rm{Q}}$. By applying deep sets operation on the last $l$ items, 
we achieve the permutation invariant property, which is consistent with local invariance~\cite{chen2019session}. Here ${\rm{Q}_1}$ is the instance-view attention query, while ${\rm{Q}_i}, i=2,3,...,L$ is the higher-level attention query. Hence, we generate $L$ queries in total, where all the queries have different receptive fields with local invariant information. 

Then, we use the generated queries,  ${\rm{Q}}_{1}$, ${\rm{Q}}_{2}$, ..., ${\rm{Q}}_{L}$ to attend to the hidden state of each item in this session. 
The multi-head attention weights are computed as 
\begin{small}
\begin{equation}
\label{con:multihead}
\begin{split}
    \alpha_{h} = \text{softmax}\left(\frac{{\mathrm{Q}}{\mathrm{W}}_h^{\mathrm{Q}}({\mathrm{K}}{\mathrm{W}}_h^{\mathrm{K}})^T}{\sqrt{d}}\right), \\
\end{split}
\end{equation}
\end{small}

\noindent where
${\mathrm{Q}}\in\mathbb{R}^{l\times d}$ is the whole query matrix, 
${\mathrm{K}}
\in\mathbb{R}^{n\times d}$ is the normalized hidden states of the items in this session, and ${\rm{W^Q_h}},{\rm{W^K_h}\in\mathbb{R}^{d\times d}}
$ are learnable parameters. $h = 1,2,...,H$ denotes the different attention head indexes. For sessions with length smaller than $L$, we consider all items in them to generate queries rather than $L$ items. 

\subsubsection{Attention Mixture to Generate Session Embeddings} 
Given $\alpha\in\mathbb{R}^{n\times lH}$, the combination of the above multi-head attention weights $\alpha_{h}$, for $ h=1,2,...,H$, to balance between capturing the most salient and the comprehensive multi-level user intent, we apply $L_{p}$ pooling ~\cite{hyvarinen2007complex} to pool the attention map and get the final session representation $\tilde{s}$ for session $S$:

\begin{equation}
\label{con:mix}
\begin{split}
\hat{\alpha}_{j,h} &=[\Sigma_{m=0}^{l-1}(\alpha_{j,mH+h})^{p}]^{\frac{1}{p}}, \\ 
s^h &= \Sigma_{j=1}^{n}\hat{\alpha}_{j,h}\tilde{h}_{j}, \\
s &= \oplus \{s^h\}_{h=1,...,H} \\
\tilde{s} &= \frac{s}{\left\|s\right\|_2}, \\
\end{split}
\end{equation}
where $\hat{\alpha}_{j,h}$ is the output of the pooling operator at location $(j,h)$,
for $j=1,2,...,n$ and $h=1,2,...,H$. $\alpha_{j,mH+h}$ is the feature value within the pooling region. $s^h$ is the session representation under the $h^{th}$ head.

Finally, we use generated session embedding to predict the last item in each session. For training we use cross-entropy loss function and Adam optimizer~\cite{kingma2014adam} to optimize the model parameters.


\subsection{Complexity Analysis}
In this section, we analyze the time complexity of Atten-Mixer to validate the effectiveness of adopting SBR-related inductive biases to prune the search space.
Given the session length as $n$, embedding dimension as $d$, we denote Atten-Mixer without adopting the SBR-related inductive biases to prune the search space as Atten-Mixer-pr. The time complexity of Atten-Mixer-pr is $O(Hn^{\lfloor \frac{n}{2} \rfloor + 2}d)$, where $H$ is the head number. After pruning the search space with SBR-related inductive biases, the time complexity would become $O(HL^{2}nd)$, where $L$ is the level number. Since $L$ is much smaller than the session length $n$, we find that the total time complexity is significantly reduced, which validates the effectiveness of our proposal.

\begin{table}[t]
\centering
  \caption{Statistics of datasets used in the experiments.}
  \label{tab:dt_descrip}
  \resizebox{.8\linewidth}{!}{
  \begin{tabular}{c c c c}
    \toprule
    Statistic & \textit{Diginetica} & \textit{Gowalla} & \textit{Last.fm} \\
    \midrule
    No. of Clicks & 981,620 & 1,122,788 & 3,835,706 \\
    No. of Sessions & 777,029 & 830,893 & 3,510,163  \\
    No. of Items  & 42,596 & 29,510 & 38,615  \\
    Average Length & 4.80 & 3.85 & 11.78  \\
    \bottomrule
  \end{tabular}
  }
\end{table}

\begin{table*}
\centering
\Large
  \caption{Results of main experiments. The results of the best performing baseline are underlined. The numbers in bold indicate statistically significant improvement (p \textless { .01}) by the pairwise t-test comparisons over the other baselines.}
  \label{tab:overall}
  \resizebox{.8\linewidth}{!}{
  \begin{tabular}{c|c c c|c c c|c c c}
    \toprule
    \multirow{2}*{Model} & \multicolumn{3}{c|}{\textit{Diginetica}} & \multicolumn{3}{c|}{\textit{Gowalla}} & \multicolumn{3}{c}{\textit{Last.fm}} \\
              ~ & HR@20 & MRR@20 & Time (s) & HR@20 & MRR@20 & Time (s) & HR@20 & MRR@20 & Time (s) \\
    \midrule
    NextItNet & $35.60$ & $9.66$ & $91.06$ & $38.69$ & $16.48$ & $67.94$ & $21.02$ & $6.46$ & $413.27$ \\
    NARM & $48.27$ & $16.43$ & $107.61$ & $49.67$ & $22.14$ & $80.52$ & $21.73$ & $6.87$ & $427.14$ \\
    SR-GNN & $51.16$ & $17.67$ & $341.68$ & $50.16$ & $24.58$ & $338.62$ & $22.49$ & $8.30$ & $1626.94$ \\
    GC-SAN & $50.63$ & $17.37$ & $437.27$ & $50.35$ & $24.65$ & $398.04$ & $22.63$ & $8.40$ & $1814.78$ \\
    SGNN-HN & $51.57$ & $17.54$ & $365.38$ & $50.72$ & $24.97$ & $326.91$ & $23.66$ & $8.34$ & $1595.59$\\
    LESSR  & $51.71$ & $18.15$ & $440.84$ & $51.34$ & $25.49$ & $511.68$ & $23.37$ & \underline{$8.84$} & $1927.20$ \\
    NISER+ & \underline{$54.18$} & $18.36$ & $292.15$ & $53.89$ & $25.73$ & $278.65$ & 23.82 & $8.36$ & $279.80$ \\
    DHCN & $53.85$ & $18.50$ & $2169.87$ & $53.77$ & $24.13$ & $2452.76$ & $22.86$ & $7.78$ & $21059.94$ \\
    DSAN & $54.02$ & \underline{$18.62$} & $273.48$ & \underline{$54.09$} & \underline{$26.64$} & $279.17$ & \underline{$24.17$} & $8.42$ & $1203.81$ \\
    \midrule
    Atten-Mixer & $\textbf{55.66}$ & $\textbf{18.96}$ & $288.12$ & $\textbf{55.12}$ & $\textbf{27.01}$ & $267.37$ & $\textbf{24.50}$ & $\textbf{9.05}$ & $1140.09$\\
    \bottomrule
  \end{tabular}
  }

\end{table*}
\section{Offline Experiments}
\label{sec:offline}
In this section, we conduct extensive experiments, and analyze the performance of the proposed Atten-Mixer model.
\subsection{Experimental Setup}
\label{sec:expe_setup}
\textbf{Dataset.} 
We evaluate the performance of Atten-Mixer and the baselines on the following three publicly available benchmark datasets, which are commonly used in the literatures of SBR~\cite{Li2017NeuralAS,Qiu2019RethinkingTI,ren2019repeatnet,Wu2019SessionbasedRW}:
\begin{itemize}
    \item \textit{Diginetica}\footnote{\noindent  http://cikm2016.cs.iupui.edu/cikm-cup}
 is a transaction dataset that is obtained from CIKM Cup 2016 Challange. Fowllowing ~\cite{Li2017NeuralAS,ren2019repeatnet,Wu2019SessionbasedRW}, we consider sessions in the last week for testing.
    \item \textit{Gowalla}\footnote{\noindent  https://snap.stanford.edu/data/loc-gowalla.html} is a dataset that contains users' check-in information for point-of-interest recommendation. Following ~\cite{guo2019streaming,tang2018personalized,chen2020handling}, we keep the 30,000 most popular locations and set the splitting interval to 1 day. We consider the last {20\%} of sessions for testing.
    \item \textit{Last.fm}\footnote{\noindent   http://ocelma.net/MusicRecommendationDataset/lastfm-1K.html} is a music-artist dataset that is used for music interest recommendation. Following ~\cite{guo2019streaming,ren2019repeatnet,chen2020handling}, we keep the 40,000 most popular artists and set the splitting interval to 8 hours. We also use the last {20\%} of sessions as the test set.
\end{itemize}
    
Following ~\cite{Wu2019SessionbasedRW,chen2020handling,guo2019streaming,tang2018personalized}, we adopt the data augmentation that has been widely applied in ~\cite{Li2017NeuralAS,Wu2019SessionbasedRW,chen2020handling} after filtering short sessions and infrequent items. Statistics of the datasets are shown in Table~\ref{tab:dt_descrip}.

\noindent \textbf{Baselines and Evaluation Metrics.}
We consider baseline models NARM\footnote{\noindent https://github.com/lijingsdu/sessionRec\_NARM}~\cite{Li2017NeuralAS}, NextItNet~\cite{yuan2019simple}, SR-GNN\footnote{\noindent https://github.com/CRIPAC-DIG/SR-GNN}~\cite{Wu2019SessionbasedRW}, GC-SAN\footnote{\noindent https://github.com/johnny12150/GC-SAN}~\cite{Xu2019GraphCS}, NISER+~\cite{gupta2019niser}, SGNN-HN~\cite{Pan2020StarGN}, LESSR\footnote{\noindent https://github.com/twchen/lessr}~\cite{chen2020handling}, DHCN\footnote{\noindent https://github.com/xiaxin1998/DHCN}~\cite{xia2021self} and DSAN\footnote{\noindent https://github.com/SamHaoYuan/DSANForAAAI2021}~\cite{yuan2021dual} to evaluate the performance of the proposed model. 

We apply grid search to find the optimal hyper-parameters for each model.
Besides, the batch size and hidden dimensionality $d$ are set to 100 and 256, respectively. We use the last 20\% of the training set as the validation set.
The ranges of other hyper-parameters are $\{1, 2, 3, ..., 10\}$ for Level-L value $L$, $\{1, 2, 4, 8, 16, 32\}$ for attention head number $H$ and $\{5\times10^{-4}, ..., 2\times10^{-2}\}$ uniformly chosen for learning rate $\eta$. We set scaled fator $\sigma$ to be 12 and $p=4$ for $L_{p}$ pooling. The optimizer we choose is Adam. We use the same evaluation metrics \textbf{HR@K} (Hit Rate) and \textbf{MRR@K} (Mean Reciprocal Rank) following previous studies ~\cite{Li2017NeuralAS,Qiu2019RethinkingTI,ren2019repeatnet,Wu2019SessionbasedRW,chen2020handling,Xu2019GraphCS,Pan2020StarGN,gupta2019niser}. All models are run five times with different random seeds and reported the average on a single NVIDIA GeForce RTX 3090 GPU.

\subsection{Overall Comparison}
\label{sec:overall_performance}
To demonstrate the overall performance of the proposed model, we compare it with the state-of-the-art recommendation methods. The experimental results of all methods are shown in Table~\ref{tab:overall}, from which we have the following observations.

GNN-based methods generally outperform RNN-based (\textit{i.e.,} NARM) and CNN-based (\textit{i.e.,} NextItNet) methods. This may because GNN has a stronger ability to explore complex graph-structured data. 

As for the Atten-Mixer model, to validate the power of the proposed Atten-Mixer operation, we simply perform it on the randomly initiated item embeddings without GNN enhanced.
As shown in Table~\ref{tab:overall}, Atten-Mixer outperforms the previous sophisticated GNN-based models even without the GNN enhanced item embeddings. 

Besides, Table~\ref{tab:overall} also shows running time per epoch of each model. Prior models such as NextItNet and NARM are incapable of maintaining satisfying performance with respect to their good efficiency. The training time of DHCN is much higher than other methods, which is about 10 times longer, revealing that propagating information along hypergraphs is quite time-consuming compared with ordinary graphs. Note that NISER+ can maintain good efficiency on \textit{Last.fm} dataset, as it only considers the last 10 items for recommendation, which may lead to severe informaiton loss when dealing with long sessions. DSAN can achieve comparable efficiency on these datasets, as it also deprecates the complex GNN designs. However, the accuracy is sacrificed due to the weak reasoning ability of the instance-level readout module. While Atten-Mixer can maintain high efficiency and achieve accurate prediction, providing the opportunity for real world applications. 

\subsection{Atten-Mixer Enhancement Study}
\label{sec:enhanced}
To figure out the universality of Atten-Mixer (\textit{i.e., how does integrating Atten-Mixer into state-of-the-art models perform compared with the original model}), we incorporate our Atten-Mixer component into two representative state-of-the-art models: SR-GNN and SGNN-HN to compare the performance difference.

\textbf{Overall performance.}
With only a few lines of code, the Atten-Mixer component can be integrated into almost any existing SBR models to aggregate the item embedding. The experimental results of several state-of-the-art methods with Atten-Mixer enhanced are illustrated in Table~\ref{tab:incor}. Atten-Mixer significantly improves the model performance with all metrics in all datasets, which 
demonstrates the universality of its application. 

\begin{table}[t]
    \centering
    \caption{Performance comparison of different session-based recommendation methods with their Atten-Mixer incorporated version. All the improvements are statistically significant at the level of p \textless { .01}.}
    \resizebox{\linewidth}{!}{
    \begin{tabular}{c|c|c c|c c|c c}
    \toprule
    \multirow{2}*{\textbf{Dataset}} & \multirow{2}*{\textbf{Metric}} & \multicolumn{2}{c|}{SRGNN} & \multicolumn{2}{c|}{SGNN-HN} & \multicolumn{2}{c}{Improv.} \\
              ~ & & w/o & w & w/o & w & SRGNN & SGNN-HN  \\

    \midrule
    \multirow{6}*{\textit{Diginetica}} & HR@5 & 26.90 & \textbf{29.67} & 24.88 & \textbf{27.68} & 10.3\% & 11.25\%\\
              ~ & HR@10 & 38.24 & \textbf{41.97} & 36.48 & \textbf{39.45} & 9.75\% &  8.14\% \\
              ~ & HR@20 & 51.16 & \textbf{55.73} & 51.57 & \textbf{52.76}  & 8.93\%  & 2.31\%\\
              ~ & MRR@5 & 15.28 & \textbf{16.58} & 13.64 & \textbf{15.48} & 8.51\%  & 13.49\%\\~ & MRR@10 & 16.78 & \textbf{18.20} & 15.17 & \textbf{17.04}  & 8.46\% & 12.33\%\\~ & MRR@20 & 17.67 & \textbf{19.16} & 17.54 & \textbf{17.97} & 8.43\%  & 2.45\% \\
    \midrule
    \multirow{6}*{\textit{Gowalla}} & HR@5 & 34.27 & \textbf{37.62} & 33.10 & \textbf{36.37} & 9.78\% & 9.88\%\\
              ~ & HR@10 & 42.18 & \textbf{46.07} & 42.98 & \textbf{44.92} & 9.22\%  & 4.51\% \\
              ~& HR@20 & 50.16 & \textbf{54.46} & 50.72 & \textbf{53.28} & 8.57\% & 5.05\%\\
              ~& MRR@5 & 22.97 & \textbf{24.46} & 21.45 & \textbf{23.27} & 6.49\% & 8.45\%   \\~& MRR@10 & 24.03 & \textbf{25.59} & 23.06 & \textbf{24.42} & 6.49\% & 5.90\%\\~& MRR@20 & 24.58 & \textbf{26.17} & 24.97 & \textbf{25.00} & 6.47\% & 0.12\% \\
    \midrule
    \multirow{6}*{\textit{Last.fm}} & HR@5 & 12.08 & \textbf{12.53} & 12.04 & \textbf{12.64} & 3.73\% & 4.98\% \\
              ~ & HR@10 & 16.57 & \textbf{17.17} & 16.72 & \textbf{17.51} & 3.62\%  & 4.72\%\\
              ~& HR@20 & 22.49 & \textbf{23.59}& 23.66 & \textbf{23.89} & 4.89\% & 0.97\% \\
              ~& MRR@5 & 7.81 & \textbf{8.22} & 7.61 & \textbf{8.06} & 5.25\% & 5.91\%  \\~& MRR@10 & 8.40 & \textbf{8.51} & 8.23 & \textbf{8.72} & 1.31\%  & 5.95\% \\~& MRR@20 & 8.80 & \textbf{8.96} & 8.84 & \textbf{9.15} & 1.82\%  & 3.51\%\\
    \bottomrule
          \end{tabular}
         }
    \label{tab:incor}
\end{table}

In addition, we can see that Atten-Mixer brings more performance improvement over original models when \textbf{K} in evaluation metrics like \textbf{HR@K} and \textbf{MRR@K} is smaller. A small value of \textbf{K} means the target items stay in the top  positions of the recommendation list. Due to the position bias ~\cite{chen2020bias} in recommendation that users tend to pay more attention to the items in a higher position of the recommendation list, our method can help original models produce more accurate and user-friendly recommendations.

\subsection{Ablation Study}
We compare Atten-Mixer to its several simplified version to investigate the contributions of multi-level user intent attention, each inductive bias as well as the attention mixture operation. For simplicity, we refer the \textit{inherent priority} as \textit{IP} and \textit{local invariance} as \textit{LI}.
The following models are tested on all datasets, where the results are reported in Table~\ref{tab:ablation}:
\begin{enumerate}
    \item (\textit{Atten-Mixer-M}) Atten-Mixer eliminates multi-level user intent attention generation and performs bottom-level attention by using the last-clicked item as the query.
    \item (\textit{Atten-Mixer-IP}) Atten-Mixer without leveraging the inherent priority, which takes all the items of the same priority along the session to aggregate information. 
    \item (\textit{Atten-Mixer-LI}) Atten-Mixer without leveraging the local invariance, which generates the higher-level user intent attention maps without using deep sets but concatenating the embeddings of last-$l$ items with linear transformation.
    \item (\textit{Atten-Mixer-LP}) Atten-Mixer without our proposed attention map pooling. We replace the $L_{p}$ pooling layer by a simple max-pooling layer over the generated attention maps.
\end{enumerate}
     
    Comparison results are presented in Table~\ref{tab:ablation}. We can notice the following observations: (1) \textit{Atten-Mixer} outperforms \textit{Atten-Mixer-M}, revealing that
compared with bottom-level reasoning, multi-level reasoning is more effective for inferring user interest.
(2) \textit{Atten-Mixer} performs better than the variants that utilize only one SBR-related inductive biases, \textit{i.e.,} \textit{Atten-Mixer-IP} and \textit{Atten-Mixer-LI}, which demonstrates that both inductive biases of SBR provide valuable information for improving recommendation accuracy.
(3) \textit{Atten-Mixer w/o Mix} performs much worse than Atten-Mixer, indicating the importance of achieving the balance between capturing the saliency and comprehensiveness of user intent in the reasoning process. However, the result is still better than most baseline models, demonstrating the importance of a reasonable readout operation with multi-level reasoning ability.

\begin{table}
\centering
\Large
  \caption{Ablation studies on different components.}
  \label{tab:ablation}
  \resizebox{\linewidth}{!}{
  \begin{tabular}{c|c c|c c|c c}
    \toprule
    \multirow{2}*{Model} & \multicolumn{2}{c|}{\textit{Diginetica}} & \multicolumn{2}{c|}{\textit{Gowalla}} & \multicolumn{2}{c}{\textit{Last.fm}} \\
              ~ & HR@20 & MRR@20  & HR@20 & MRR@20 & HR@20 & MRR@20 \\
    \midrule
    
    Atten-Mixer-M & 52.27 & 17.52 & 50.20 & 25.06 & 22.14 & 8.30 \\
    Atten-Mixer-IP & 53.76 & 17.99 & 52.62 & 25.03 & 22.56 & 8.83 \\
    Atten-Mixer-LI & 53.81 & 17.78 & 52.27 & 24.91 & 22.28 & 8.81 \\
    Atten-Mixer-LP & 53.48 & 17.89 & 53.29 & 25.54 & 23.44 & 8.90 \\
    \midrule
    Atten-Mixer & \textbf{55.66} & \textbf{18.96} & \textbf{55.12} & \textbf{27.01} & \textbf{24.50} & \textbf{9.05} \\
    \bottomrule
  \end{tabular}
  }
\end{table}


\subsection{Hyper-parameter Sensitivity Analysis}
We study how the value of $l$ in Level-L and the number of heads affect the performance of the proposed method. 

\begin{figure}[h]
\centering
  \subfigure[\textit{Diginetica}]{\includegraphics[width=.32\linewidth]{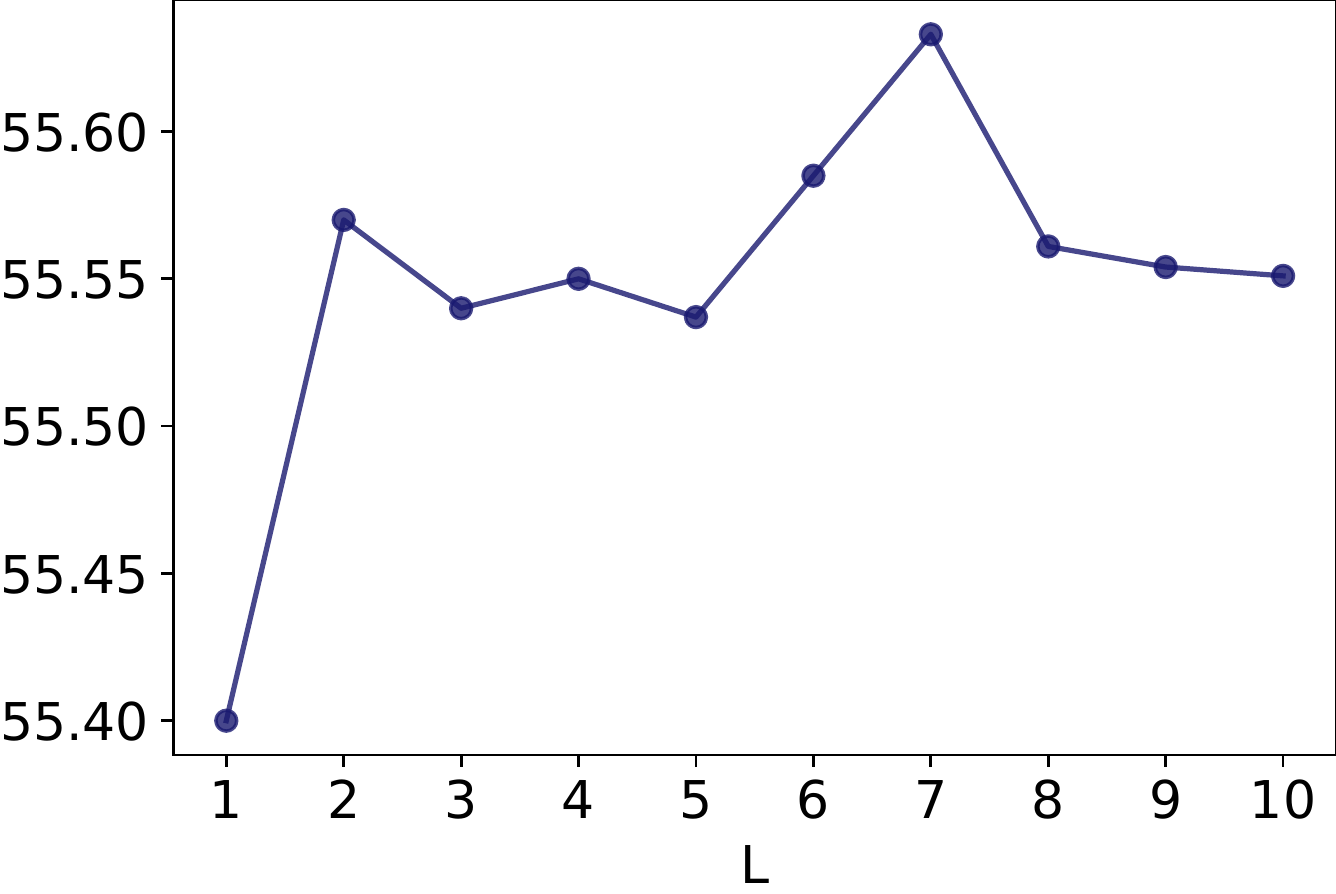}}
    \subfigure[\textit{Gowalla}]{\includegraphics[width=.31\linewidth]{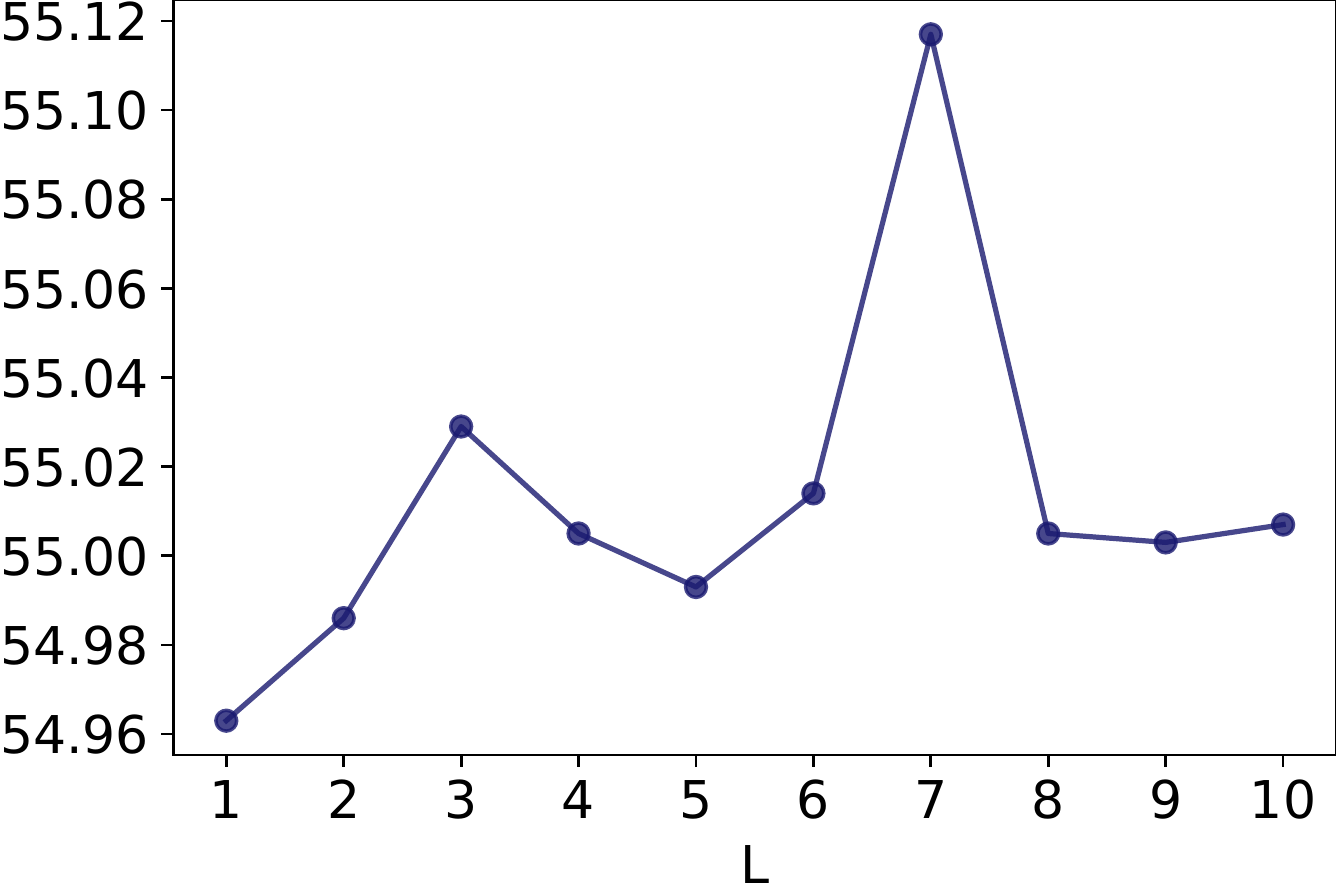}}
      \subfigure[\textit{Last.fm}]{\includegraphics[width=.31\linewidth]{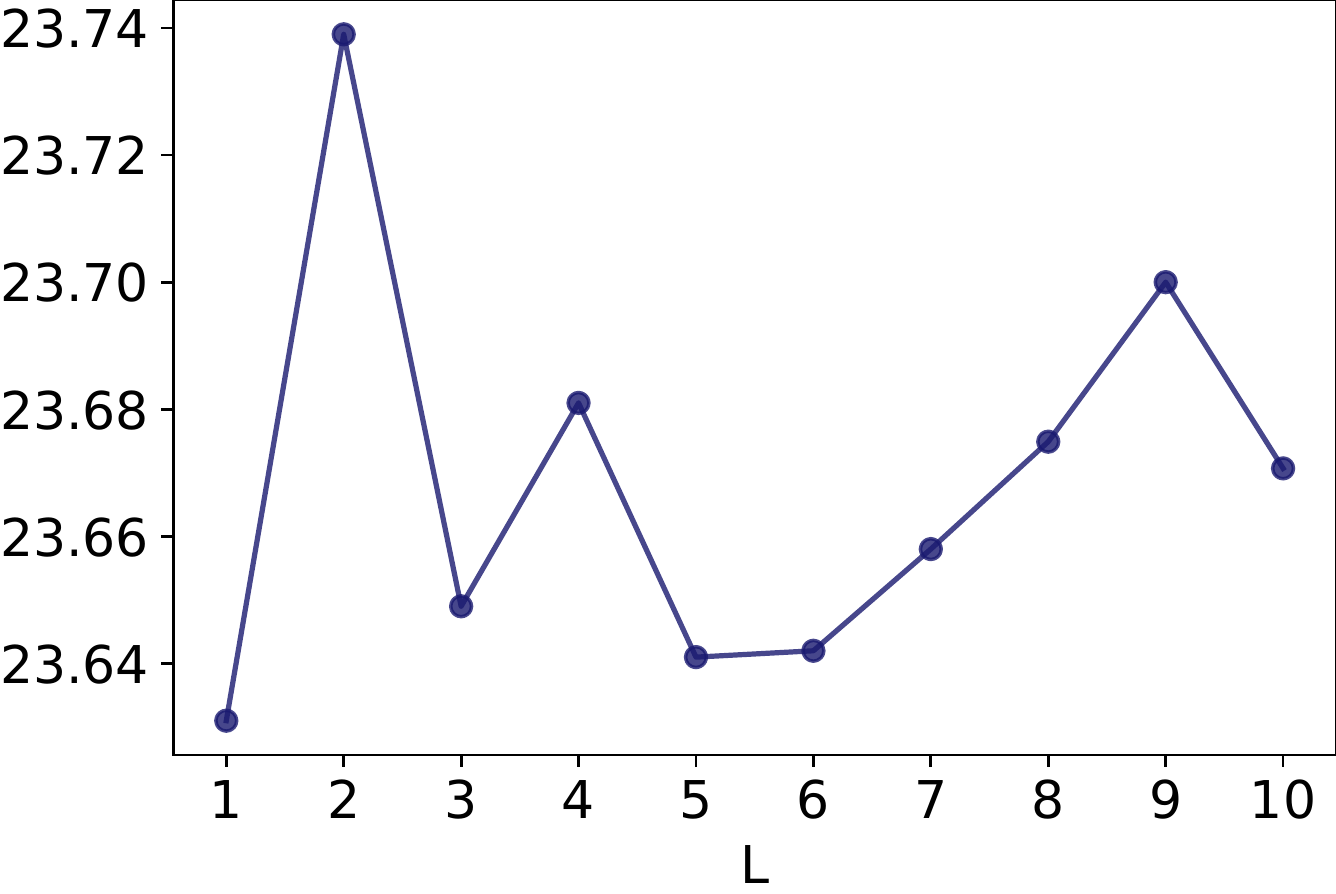}}
    \caption{HR@20 w.r.t. the value of L.}
    \label{fig:ablation_k_hr}
\end{figure}

\noindent\textbf{Impact of L.} We consider changing the $l$ to study the influence of different $l$ values. The results are shown in Figure~\ref{fig:ablation_k_hr}. We have the following observations: (1) Atten-Mixer performs worst when $l = 1$. This is because when $l = 1$, we generate the overall preference with attention related to the last-clicked item. Without the reasoning process brought by the GNN propagation part, simply relying on the instance-view readout module is fragmented as it lacks the information about high-level connections between items. 
(2) At the beginning, increasing the number of $l$ significantly improves the model performance, which demonstrates the importance of multi-level user intent reasoning over item transitions.
(3) The performance reaches the peak when $l$ is maintained in a certain range (\textit{eg.} $l=2$ on \textit{Last.fm}), and starts to decrease after we continue to increase $l$. As $l$ keeps increasing, the inherent priority of emphasizing the last several items will be degraded, which may create uninformative higher-level user intent for our attention component.

\begin{figure}[h]
\centering
  \subfigure[\textit{Diginetica}]{\includegraphics[width=.32\linewidth]{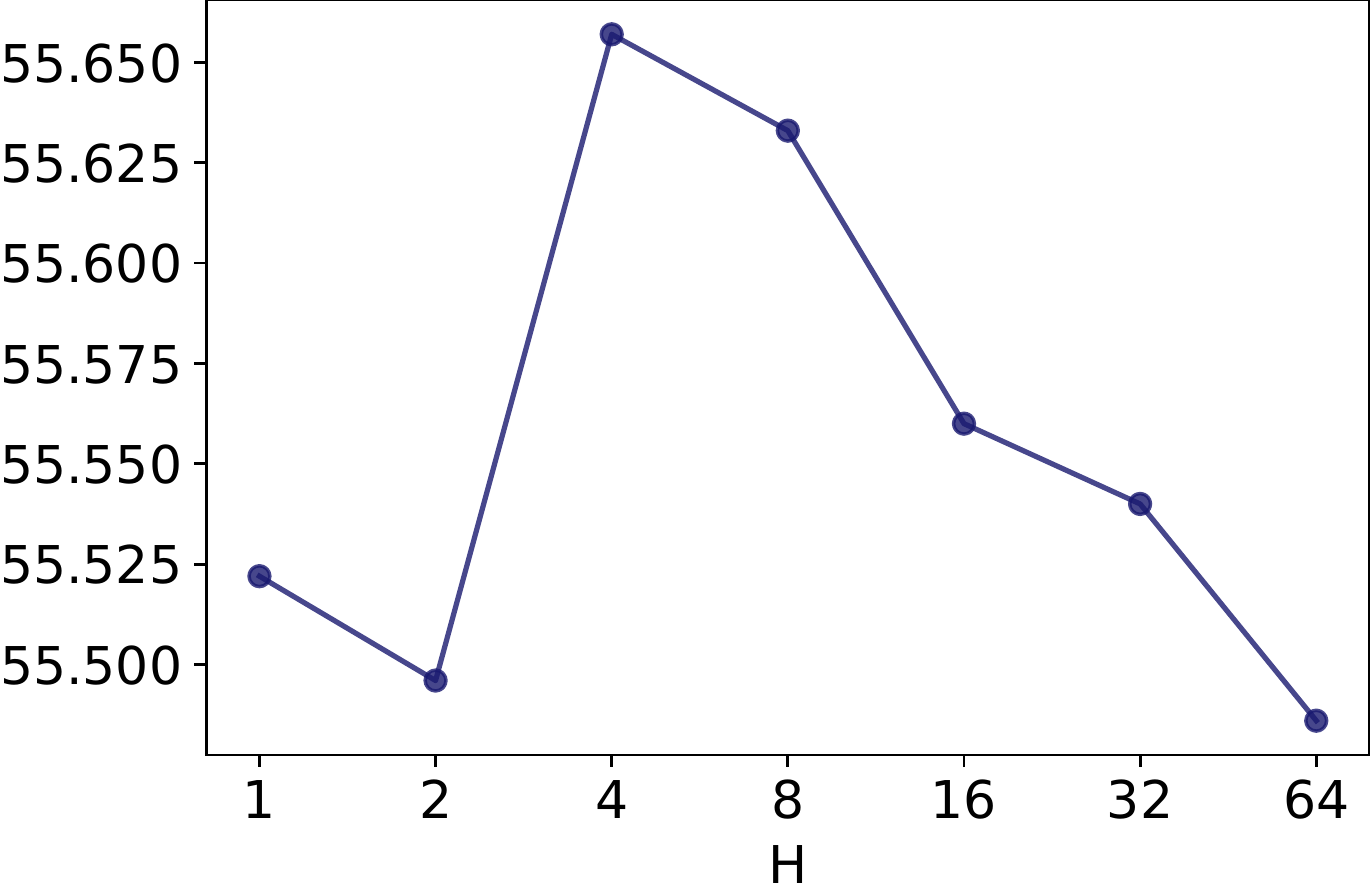}}
  \subfigure[\textit{Gowalla}]{\includegraphics[width=.31\linewidth]{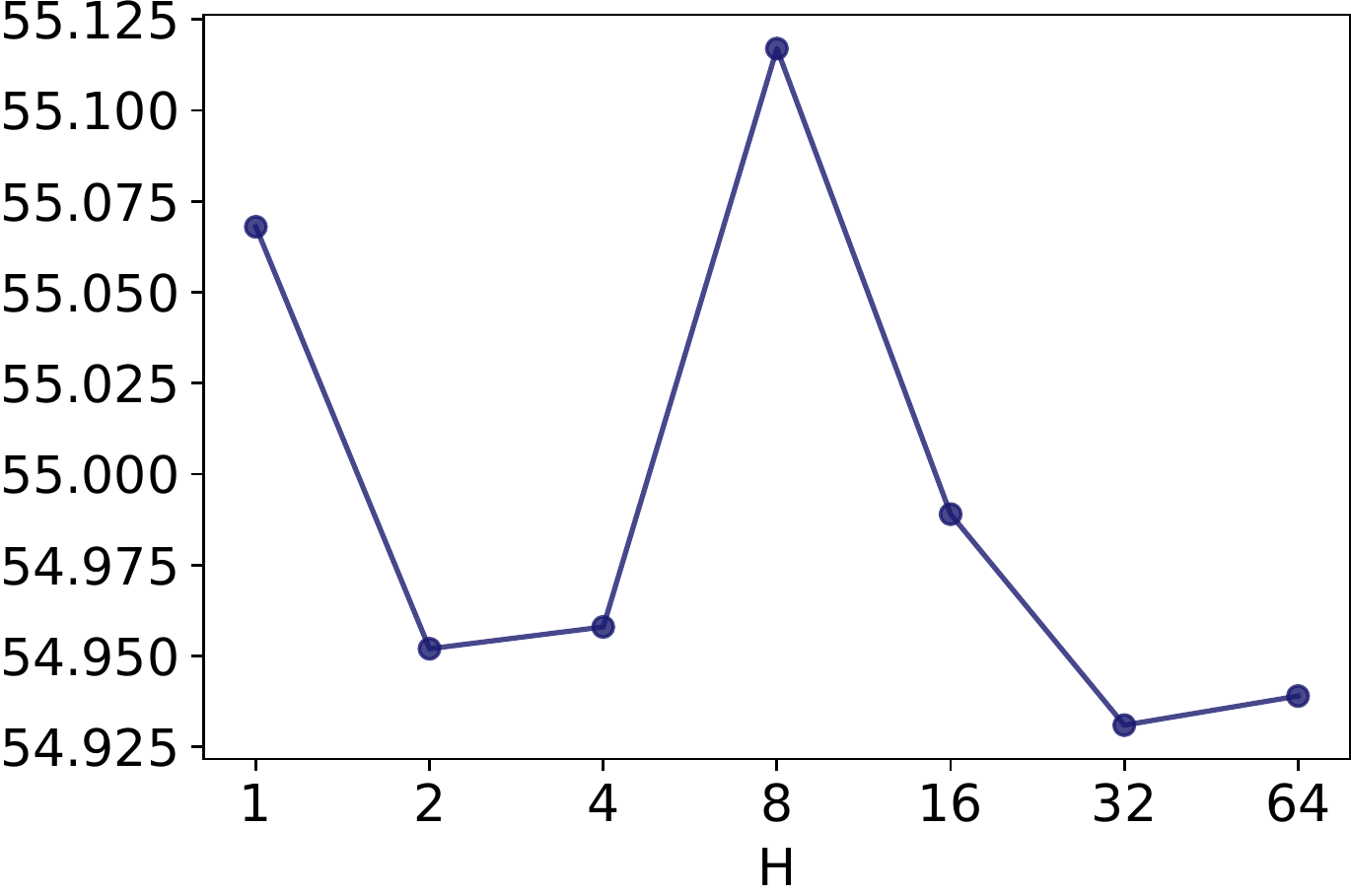}}
  \subfigure[\textit{Last.fm}]{\includegraphics[width=.31\linewidth]{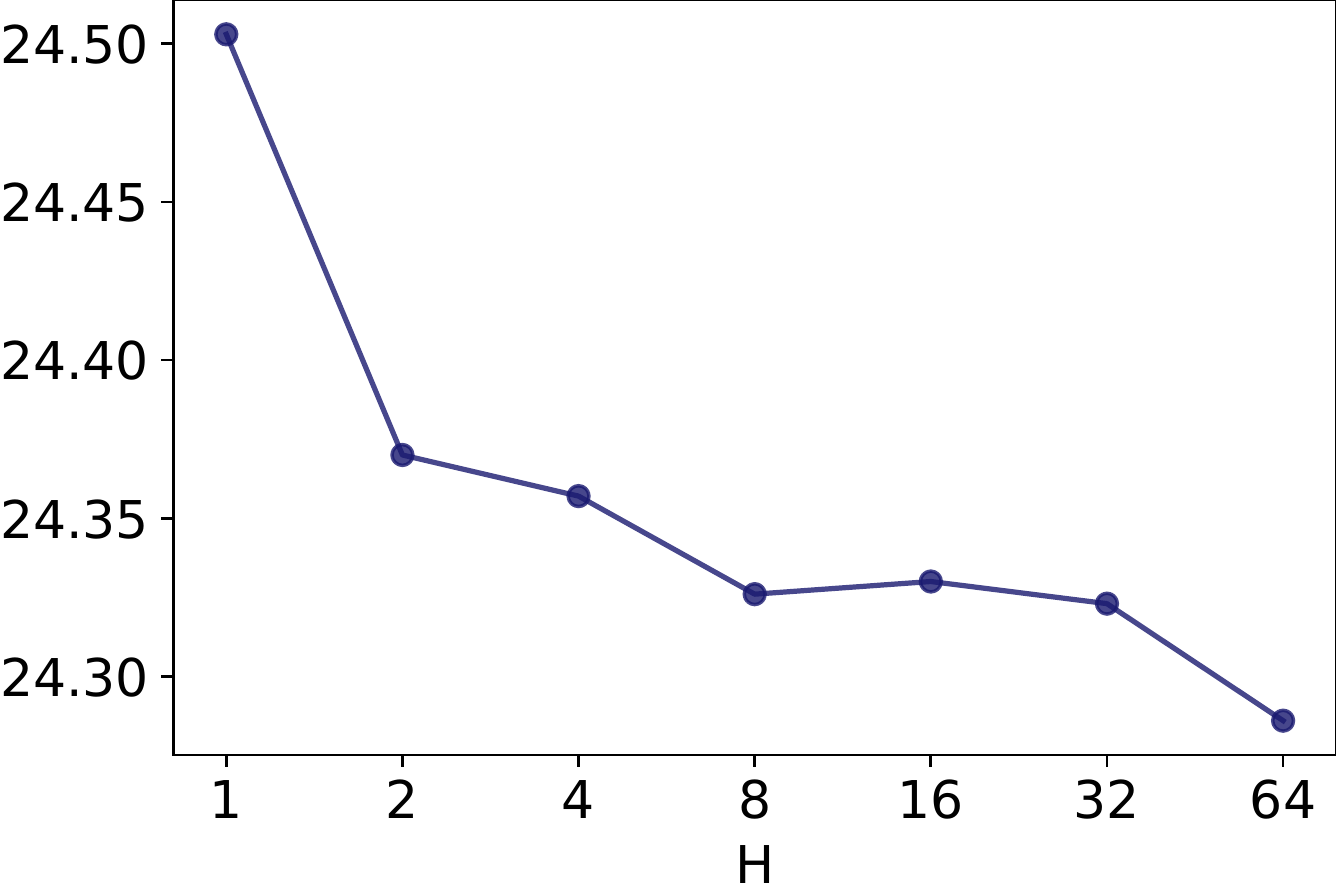}}
    \caption{HR@20 w.r.t. the value of H.} 
    \label{fig:ablation_h_hr}
\end{figure}

\noindent\textbf{Impact of Multi-Heads.}
To study the impact of $H$, we test Atten-Mixer with different settings of $H$. According to Figure~\ref{fig:ablation_h_hr}, on \textit{Diginetica} and \textit{Gowalla} datasets, the performance increases with the increment of $H$. That is reasonable because a larger $H$ means model can comprehensively utilize more aspects of information for the target items, which provides a more precise characterization of the item sequence’s potential targets. On \textit{Last.fm} dataset, the performance reaches its best when the head equals 1, indicating that more heads gather noises due to the larger variance of this dataset.

\begin{figure}[htbp]
  \begin{center}
  \subfigure[HR@20 on \textit{Diginetica}]{\includegraphics[width=.47\linewidth]{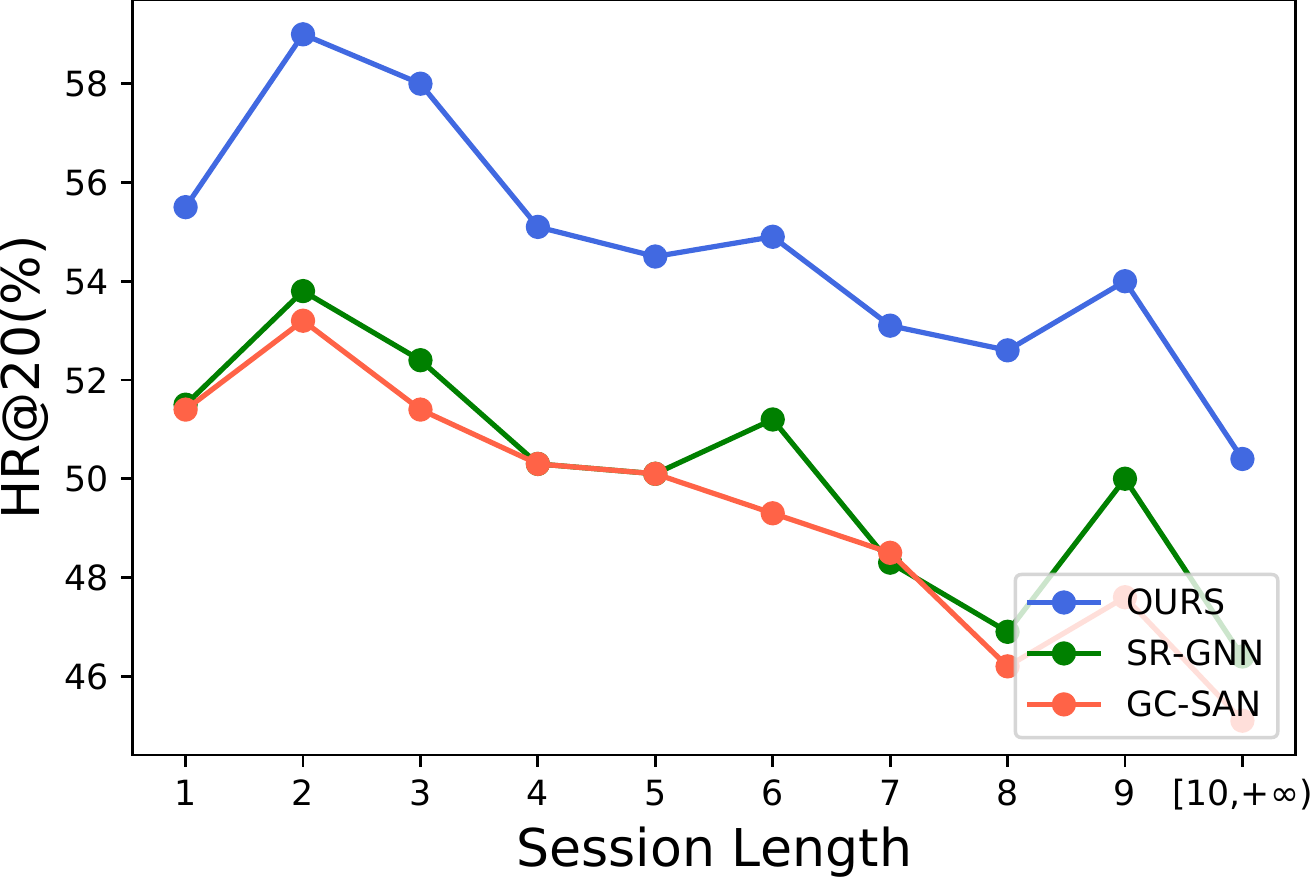}\label{fig:hd}}
  \subfigure[MRR@20 on \textit{Diginetica}]{\includegraphics[width=.47\linewidth]{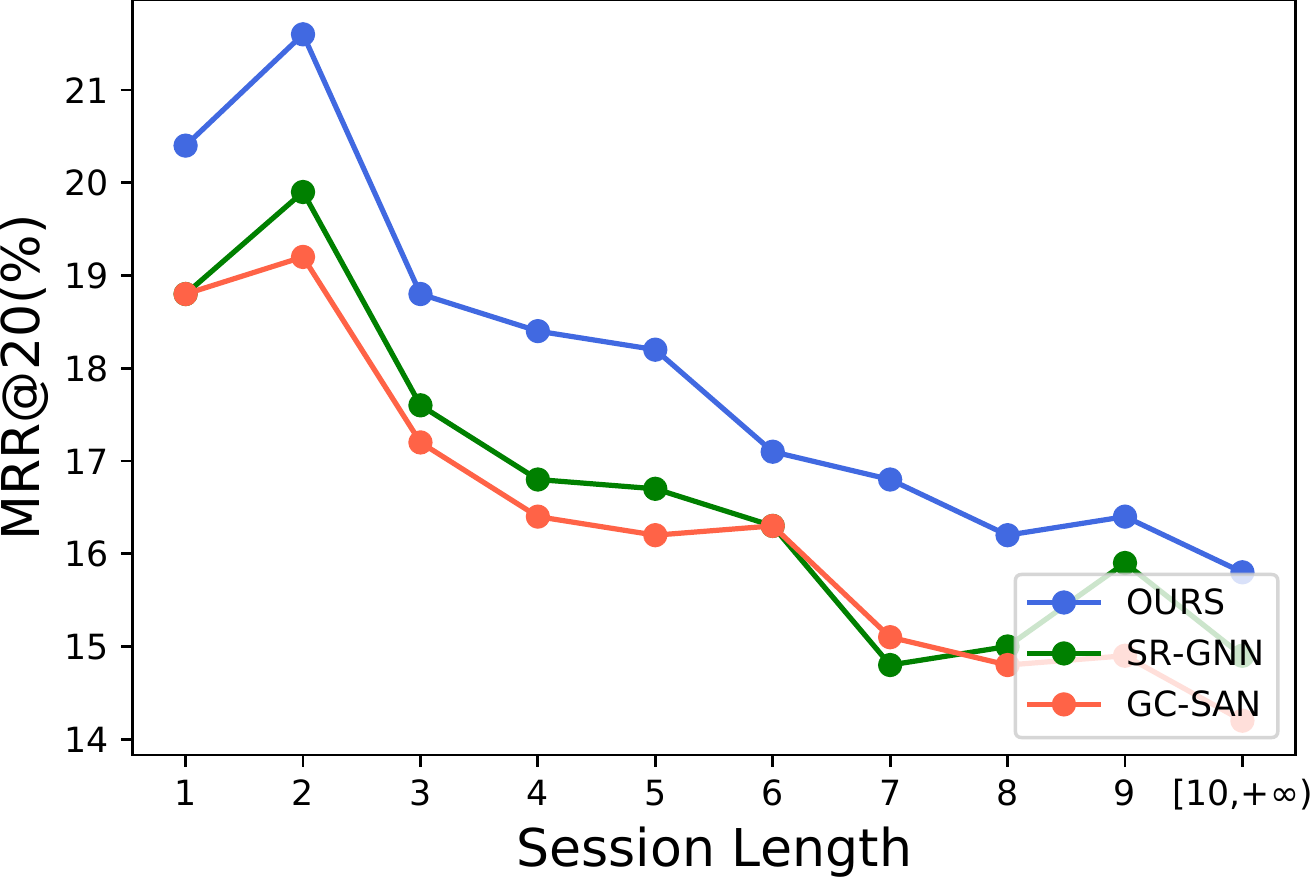}\label{fig:md}}

  \end{center}
  \caption{Performance w.r.t. different session length.} 
  \label{fig:session_length}
\end{figure}

\subsection{Impact of Different Session Lengths}
To figure out how Atten-Mixer performs on sessions with different lengths compared to the existing readout components, we evaluate GNN with Atten-Mixer and GNN with previous readout operations, including SR-GNN and GC-SAN
on \textit{Diginetica} dataset.

From Figure~\ref{fig:session_length}, we notice that as the session length increases, the performance of all models on \textit{Diginetica} dataset consistently decreases, which may be because longer sessions are more likely to contain unrelated items, making it harder to identify the user preference correctly. GNN with Atten-Mixer achieves the best performance. We attribute the difference in performance between GC-SAN and GNN with Atten-Mixer to: (1) compared with SR-GNN, which only emphasizes the last clicked item, the GNN with Atten-Mixer makes the information from long-range items available in information propagating, which can effectively alleviate the information loss problem; and (2) compared with GC-SAN which uses the general attention without item priority, the attention mixer component in GNN with Atten-Mixer allows for the various priorities within the sessions to be investigated more accurately, which boosts the ranking of the target item in the recommendation list.

\begin{figure}[h]
    \centering
    \includegraphics[width=.6\linewidth]{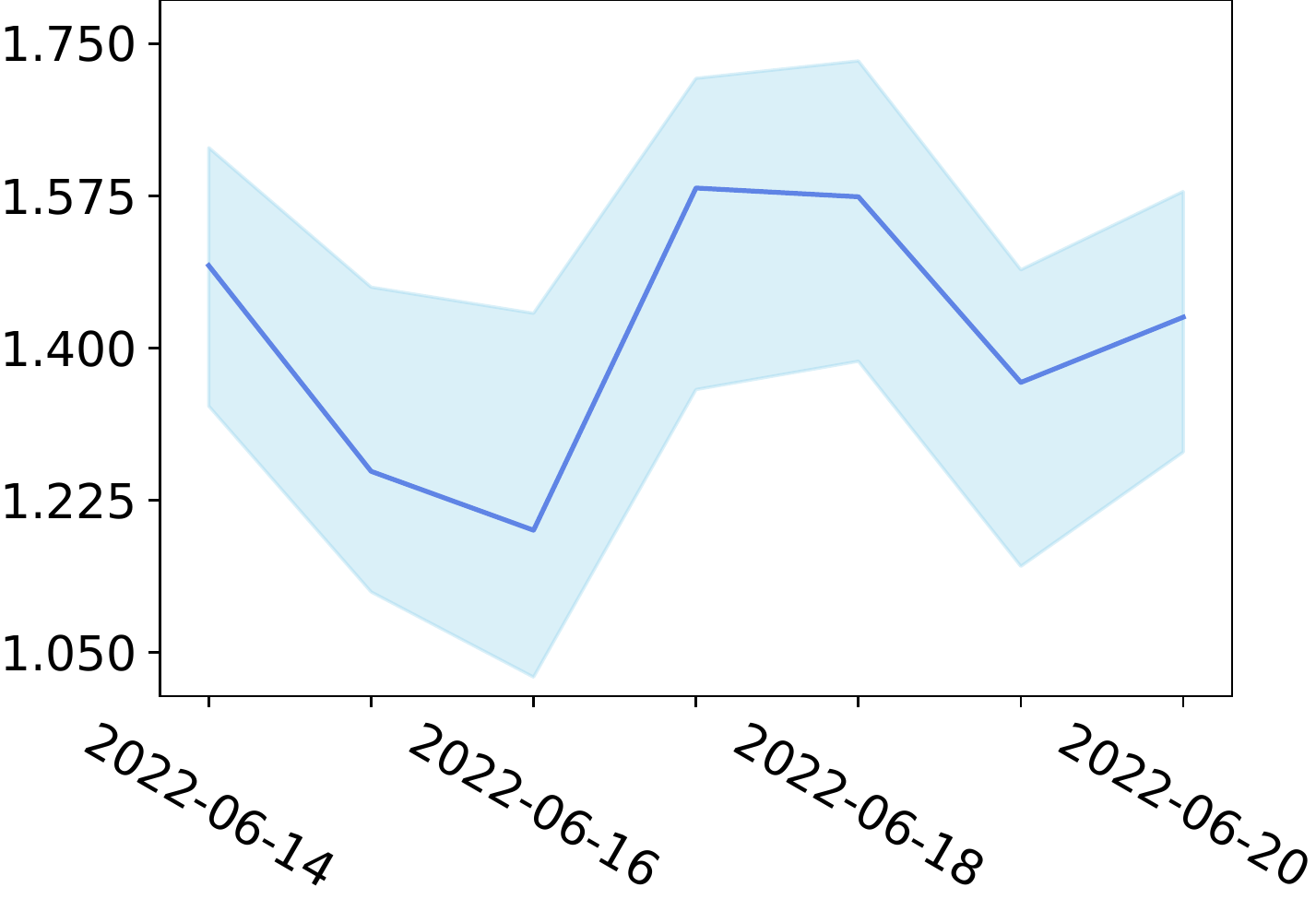}
    \caption{Top business metric improvement percentage (y-axis) over days (x-axis) in online experiments.}
    \label{fig:online}
\end{figure}
\section{Online Performance Analysis}
Atten-Mixer has been fully depolyed into several heavy-traffic scenarios since April 2021. For a natural extension of the offline experiments conducted in Section~\ref{sec:offline}, we conduct a week's online experiment in real-world settings. We add the Atten-Mixer on top of the existing well-tuned SR-GNN model used in production. Figure~\ref{fig:online} shows our main online results conducted in these heavy-traffic scenarios, with 
millions of page views each day. In a time frame of 7 days, Atten-Mixer has been consistently outperforming the previous baseline model, with +1.5\% increase in top business metrics. Given the results, the Atten-Mixer empowered model has been successfully launched in the production system.


\section{Conclusion}
We investigate the classical GNN-based SBR models, and discover that they are over-parameterized, based on which we proposed Atten-Mixer, an efficient and effective SBR model with multi-level reasoning component through reflecting on the human reasoning process. Extensive online and offline analyses validate our proposal. 


\newpage
\bibliographystyle{ACM-Reference-Format}
\balance
\bibliography{sample-base}

\appendix

\end{document}